\documentclass[english]{article}
\usepackage[T1]{fontenc}
\usepackage[latin9]{inputenc}
\usepackage{amsmath}
\usepackage{amssymb}
\usepackage{graphicx}
\usepackage{esint}
\usepackage{color}

\makeatletter
\newcommand{\lyxaddress}[1]{
\par {\raggedright #1
\vspace{1.4em}
\noindent\par}
}

\usepackage{babel}

\makeatother

\usepackage{babel}

\begin{document}

\title{Perturbative Semiclassical Trace Formulae for Harmonic Oscillators}

\author{J M{\o}ller-Andersen$^{1}$ and M \"{O}gren$^{1,2,3}$}

%PACS
% 03.65.Sq Semiclassical theories and applications
%
% 11.30.Qc Spontaneous and radiative symmetry breaking
% 
%MSC
% 81Q15 Perturbation theories for operators and differential equations
%
% 81Q20 Semiclassical techniques including WKB and Maslov methods
%
% 81R40 Symmetry breaking

\date{\today{}}

\maketitle

\lyxaddress{$^{1}$Department of Applied Mathematics and Computer Science, Technical
University of Denmark, 2800 Kgs. Lyngby, Denmark.\\
 $^{2}$Nano Science Center, Department of Chemistry, University of
Copenhagen, Universitetsparken 5, 2100 K{\o}benhavn {\O}, Denmark.\\
 $^{3}$School of Science and Technology, \"{O}rebro University, 70182 \"{O}rebro, Sweden.\\
 (e-mails: jakmo@dtu.dk; magnus@ogren.se)}

\begin{abstract}
In this article we extend previous semiclassical studies by including more general perturbative potentials of the harmonic oscillator in arbitrary spatial dimensions.
Our starting point is a radial harmonic potential with an arbitrary even monomial perturbation, which we use to study the resulting $\mathrm{U}(D)$ to $\mathrm{O}(D)$ symmetry breaking.
We derive the gross structure of the semiclassical spectrum from periodic orbit theory, in the form of a perturbative ($\hbar \rightarrow 0$) trace formula.
We then show how to apply the results to even order polynomial potentials, possibly including mean-field terms.
We have drawn the conclusion that the gross structure of the quantum spectrum is determined from only classical circular- and diameter-orbits for this class of systems.

\end{abstract}

%%\keywords{perturbative trace formula, semiclassical density of states, radially perturbed harmonic oscillators.}

\section{Introduction}

In 1913 Niels Bohr published his seminal work on the Hydrogen atom
\cite{Bohr1913} where he depicted the electron orbiting the proton
as planets orbits the sun. Bohr's pictorial model is used in logotypes
of research institutions and companies world wide, and is still the
most popular way to draw an atom. With the vocabulary of today, Bohr
obtained the quantum mechanical energy levels of the Coulomb potential.
But 100 years ago, there was hardly any established quantum theory.
Bohr's way of combining well known classical mechanical laws with
an innovative quantization of the electron radius (or equivalently,
its angular momenta), together with the correspondence principle,
had a tremendous influence on the development taking place in the following two
decades. Bohr only included circular orbits, although
the corresponding classical system have elliptic solutions \cite{Sommerfeld1916}.
It must be considered of great historical importance for the development
of quantum mechanics that Bohr obtained the correct quantum spectrum
from his simple model. At first sight, the development of the Heisenberg-Schr\"{o}dinger
quantum theory seemed to be unrelated to Bohr's semiclassical treatment.
However, work by Einstein \cite{e}, Brillouin \cite{b} and Keller \cite{k} (EBK), Van Vleck \cite{VanVleck1928},
and Feynman \cite{Feynman1948}, have pointed onto relations between the
action for the classical orbits of a particle and the corresponding quantum spectrum.
The interest in relations between classical systems
and their quantum counterpart boosted again in the 60s and 70s, due to new powerful computers, with
the study of quantum chaos \cite{chaos-book1, chaos-book2, chaos-book3}. 
Finally a so called periodic orbit theory
(POT) was introduced for chaotic systems by Gutzwiller \cite{gutz}, and for different regular systems by Balian and Bloch \cite{bablo}, and Berry, Mount and Tabor \cite{bermo, bertab}, and others.
Now the studies of
the periodic classical orbits were related to quantum mechanical observables
through so called trace formulae (TF), which had already been studied
in the 50s by Selberg \cite{Selberg1956}. 
A trace formula expresses the spectrum of a differential operator, as for example in the time independent Schr\"{o}dinger equation,  represented by a
train of delta functions, with a sum over the classical periodic orbits. 
Several principally important quantum systems:
the harmonic oscillator; the cavity; and again, the hydrogen atom,
were soon analyzed within POT \cite{bablo,BBBook}. 
These systems could all be connected by the principle trace formula for integrable systems given by Berry and Tabor \cite{bertab}.
At the same time, semiclassical
approximations were succesful in describing shell structures of different
quantum many-body systems from atomic and nuclear physics \cite{struma, struma2}.
This was one motivation to also refine the POT further to classical chaotic systems.
Studies of (super-) shell structures have now  been undertaken in new man-made
systems: the abundance in atomic metal clusters \cite{nish}, that
have been confirmed experimentally \cite{klavs}; in solid state devices,
like the conductance of quantum wires \cite{wire, wire2}; in weakly repulsive
atomic Fermi gasses \cite{YuPRA2005}, and pairing gaps of attractive
Fermi gasses \cite{OgrenPRA2007}, nano-grains \cite{OlofssonPRL2008} and recently also in triangular
flakes of graphene \cite{AkolaPRB2008}.

In this article we treat a class of quantum systems of principal importance in approximations,
the isotropic perturbed harmonic oscillator (HO) in arbitrary dimensions.
We here present a TF for $\mathrm{U}\left(D\right)$ to $\mathrm{O}\left(D\right)$
symmetry breaking that gives the gross structure of density of states (DOS), sometimes
called the level density, to leading order in $\hbar^{-1}$ for the
perturbed system and that recovers the quantum mechanical TF
in the limit of no perturbation to leading order in $\hbar^{-1}$. 
The breaking of $\mathrm{U}\left(D\right)$ symmetry for collections of HOs can have future relevance for many different applications, from nuclear physics and clusters, to more recent systems with cold quantum gasses and graphene. 
Special cases have
been presented before: the quartically perturbed two-dimensional HO
was treated by Creagh in \cite{crpert}; and the three-dimensional
counterpart by Brack \textit{et al.} in \cite{JoPA2005}. 
The present treatment follows
a similar perturbative technique as pioneered by Creagh \cite{crpert},
but generalize the special case of quartic perturbation and also allows the treatment
of arbitrary dimensions simultaneously.

\section{The $D$-dimensional Harmonic Oscillator}

We consider the Hamiltonian of the $D$-dimensional harmonic oscillator
(HO), as given by the following Hamiltonian function defined from
the classical space and momentum coordinates $\mathbf{q},\mathbf{p} \in \mathbb{R}^D$ 
\begin{equation}
H_{0}\left(\mathbf{q},\mathbf{p}\right)=\sum_{j=1}^{D}\left(\frac{p_{j}^{2}}{2m}+\frac{1}{2}m\omega_{j}^{2}q_{j}^{2}\right).\label{eq: HarmonicHamiltonian}
\end{equation}
We consider an isotropic HO, $\omega\equiv\omega_{j},\ j=1,2,...,D$,
with unit mass, $m=1$, such that the characteristic length scale
of the oscillator is $R_{0}=\sqrt{2E}/\omega$.

\subsection{Classical mechanics of the Harmonic Oscillator}
\label{sec: ClassicalHO}

From the Hamiltonian (\ref{eq: HarmonicHamiltonian}) we can
deduce Hamilton's equations, with solutions
\begin{equation}
\left\{ \begin{array}{l}
\dot{\mathbf{q}}\left(t\right)=\mathbf{p}\left(t\right)\\
\dot{\mathbf{p}}\left(t\right)=-\omega^{2}\mathbf{q}\left(t\right)
\end{array}\right.,\quad\left\{ \begin{array}{l}
\mathbf{q}\left(t\right)=\mathbf{q}_{0}\cos\left(\omega t\right)+\frac{\mathbf{p}_{0}}{\omega}\sin\left(\omega t\right)\\
\mathbf{p}\left(t\right)=\mathbf{p}_{0}\cos\left(\omega t\right)-\omega\mathbf{q}_{0}\sin\left(\omega t\right)
\end{array}\right.,\label{eq:HamiltonsEquationAndSolution}
\end{equation}
where $\mathbf{q}_0,\mathbf{p}_0$ are constant vectors. The solutions are circles in some two-dimensional hyper-plane of the $\left( \mathbf{q},\mathbf{p} \right)$ phase-space, which projects to ellipses
in the $D$-dimensional ($ \mathbf{q}$-) configuration space. Note that many different choices of $\mathbf{p}_0,\mathbf{q}_0$ give rise to the same orbit. For a constant energy
$E=H_{0}\left(\mathbf{q}_0,\mathbf{p}_0\right)$ we have from (\ref{eq: HarmonicHamiltonian})
and (\ref{eq:HamiltonsEquationAndSolution})
\begin{equation}
H_{0}\left(\mathbf{q}(t),\mathbf{p}(t)\right)=\frac{1}{2}\left(\left|\mathbf{p}_{0}\right|^{2}+\omega^{2}\left|\mathbf{q}_{0}\right|^{2}\right)=E.\label{eq:EnergySurface}
\end{equation}
That is, energy is conserved along orbits. Consequently the normalised solutions $\left(\mathbf{q}(t)/R_{0},\,\mathbf{p}(t)/\omega R_{0}\right)$
live on the unit sphere $S^{2D-1}$ in phase-space. 

Identifying $\mathbb{R}^{2D} \simeq \mathbb{C}^D$ by $\mathbf{z} \sim (\mathbf{q},\mathbf{p})$ with $\mathbf{q} = \text{Re}(\mathbf{z})/\omega$ and $\mathbf{p} = - \text{Im}(\mathbf{z})$, 
the Hamiltonian (\ref{eq: HarmonicHamiltonian}) can be rewritten
to 
\begin{equation}
H_{0}(\mathbf{z})=\frac{1}{2}\overline{\mathbf{z}}^{T}\mathbf{z}.\label{eq: HarmonicHamiltonianComplex}
\end{equation}
Here $\mathbf{z}^{T}$ denotes transposition of the (column) vector
$\mathbf{z}\in\mathbb{C}^{D}$, and a bar means complex conjugation.
It then directly follows that the system has $\mathrm{U}(D)$-symmetry (invariant under the action of a $D$ dimensional unitary matrix), since
given a matrix $\mathbf{A}\in \mathrm{U}(D)$ we obtain from (\ref{eq: HarmonicHamiltonianComplex})
\begin{equation}
H_{0}\left(\mathbf{A}\mathbf{z}\right)=\frac{1}{2}\overline{\mathbf{A}\mathbf{z}}^{T}\mathbf{A}\mathbf{z}=\frac{1}{2}\overline{\mathbf{z}}^{T}\overline{\mathbf{A}}^{T}\mathbf{A}\mathbf{z}=H_{0}\left(\mathbf{z}\right).
\label{U_D_Matrix_Symmetry}
\end{equation}
Now Hamilton's equation and solution (\ref{eq:HamiltonsEquationAndSolution})
simply reads
\begin{equation}
\dot{{\mathbf{z}}}=i \omega \mathbf{z},\quad\mathbf{z}(t)=e^{i\omega t}\mathbf{z}_{0}.\label{eq: U1Solution}
\end{equation}
The formulation of the orbits in the phase-space are then simply
\begin{equation}
\mathbf{q}(t)=  \text{Re}\left\{ e^{i\omega t}\mathbf{z}_0 \right\} / \omega ,\:\mathbf{p}(t)= - \text{Im}\left\{ e^{i\omega t}\mathbf{z}_0 \right\} ,
\end{equation}
and an alternative real parametrisation
to \eqref{eq:HamiltonsEquationAndSolution} is 
\begin{equation}
\left\{ \begin{array}{l}
\mathbf{q}(t)=\mathbf{R_{0}}\cos\left(\omega t+\mathbf{\nu}\right),\:\nu_{1},...,\nu_{D}\in\left[0,\:2\pi\right)\\
\mathbf{p}(t)=\dot{\mathbf{q}}(t)
\end{array}\right.,\label{eq:AlternativeParametrisation}
\end{equation}
where $\mathbf{R_{0}}=\left(\sqrt{2E_{1}}/\omega,\,...,\,\sqrt{2E_{D}}/\omega\right)$
with $E_{1}+...+E_{D}=E$. 
We now choose the initial time (e.g.) by setting $\nu_{1}=0$, such that we determine the initial values of the first components of the phase-space coordinates to be $\mathbf{q}_1(0)=\sqrt{2E_1}/\omega $ and $\mathbf{p}_1(0)=0$.
Then the constant vector $\mathbf{z}_0/ \omega R_{0}$ can be viewed as living in the complex projective space $\left(n_{1},n_{2}e^{i\nu_{2}},...,n_{D}e^{i\nu_{D}}\right)\in\mathbb{C}P^{D-1}$ \cite{BengtssonIJMPA2002}. 
Here the $D-1$ complex parameters in  $\mathbb{C}P^{D-1}$ corresponds to $D-1$ real angles that
parametrise part of $S^{D-1}$, i.e., for $R_{0}>0$ we have $n_{1},...,n_{D}\in\left[0,\:1\right]$,
with $n_{1}^2+...+n_{D}^{2}=1$ due to the energy conservation, together with the $D-1$ phase angles
$\nu_{2},...,\nu_{D}\in\left[0,\:2\pi\right)$ remaining free when $\nu_{1}=0$. 
This explains the background for the two possible alternative calculations outlined in \cite{JoPA2005} for $D=3$.

As mentioned earlier, many different choices of $(\mathbf{q}_0,\mathbf{p}_0)$ leads to the same orbits. 
As we will see the high dimensional symmetry allows the short mathematical description of all orbits of the same energy. 
The discussion below will be short and informal, as the details of the spaces and identifications we mention are covered in standard literature on symplectic geometry and classical mechanics, see for example \cite{ArnoldClassicalMechanics}. 
According to \eqref{eq: U1Solution}, $\mathrm{U}(1)$ acts on solutions $\mathbf{z}(t)$
by time. The remaining symmetry is hence $\mathrm{SU}(D)\simeq \mathrm{U}(D)/\mathrm{U}(1)$, corresponding to the space of "special" unitary matrices of determinant one. 
As \eqref{eq: U1Solution} also shows; an orbit is completely contained in some complex "line" (a real two-dimensional hyper-plane), which we without loss of generality might assume to be the line spanned by the first complex coordinate. 
The group of matrices which fixes the first coordinate of a vector while preserving the energy is $\mathrm{U}(D-1)$.
Removing this symmetry finally gives us $\mathbb{C}P^{D-1}\simeq \mathrm{SU}(D)/\mathrm{U}(D-1)$. So the space of all solutions of the same energy can indeed be parametrized by the complex projective space,
in agreement with the specific parametrisation (\ref{eq:AlternativeParametrisation})
with $\nu_{1}$ fixed. A dimension count shows that this yields exactly all of the solutions. 

Finally, another way to describe this manifold of solutions, which will be of particular use for us, is the following:
 \eqref{eq: U1Solution} shows that $S^1$ acts on the energy sphere $S^{2D-1}$ ($H_0 = E$); the well known quotient space $S^{2D-1}/S^1 \simeq \mathbb{C}P^{D-1}$ is realized by the famous Hopf map, which end up being a so-called Riemannian submersion when equipping $\mathbb{C}P^{D-1}$ with the Fubini-Study (FS) metric \cite{Sakai}. Hence schematically it reads
\begin{equation}
(\mathbb{R}^{2D},g_{\mathbb{R}})\xrightarrow{H_{0}=E}(S^{2D-1},g_{\mathrm{can}})\xrightarrow{(S_{1},\pi_{\mathrm{Hopf}})}(\mathbb{C}P^{D-1},g_{\mathrm{FS}}).
\end{equation}
In local coordinates, this allows us to write the volume measure on $S^{2D-1}$ as
\begin{equation}
d \mathrm{vol}_{S^{2D-1}} = d \mathrm{vol}_{\mathbb{C}P^{D-1}} d t
\end{equation}
which will be used in sections \ref{sec: PertubativeTraceFormula} and \ref{sec: ModulationFactor}. Here $d \mathrm{vol}$ is the Riemannian volume form: the canonical choice of volume measure induced by the metric.

\subsection{Trace formula for the HO}

The well known quantum mechanical energy spectrum of the $D$-dimensional harmonic
oscillator is
\begin{equation}
E_{n}=\hbar\omega\left(n+D/2\right),\: n=0,1,2,...,\label{eq:HOQME}
\end{equation}
where each energy has a degeneracy factor
\begin{equation}
d_{n}= \binom {n+D-1} {D-1} = \frac{1}{\left(D-1\right)!}\prod_{j=1}^{D-1}\left(n+j\right).\label{eq:HOQMd_n}
\end{equation}
An energy spectrum can be expressed in the form of a trace formula for the density of states (DOS) \cite{gutz,Selberg1956,BBBook}
\begin{equation}
g\left(E\right)\equiv \bar{g} + \delta g = g_{ETF}(E) +\sum_{\gamma}{\cal {A}}_{\gamma}(E)\cos\left(\frac{S_{\gamma}(E)}{\hbar}-\mu_{\gamma}\frac{\pi}{2}\right).\label{eq:GeneralTF}
\end{equation}
 The first term in (\ref{eq:GeneralTF}) $\bar{g}$, is the extended Thomas-Fermi
DOS \cite{BBBook,Ogren}, which is a smoothly varying function of energy.
The second term, being built up by the summation over classical periodic
orbits $\gamma$ with amplitudes ${\cal {A}}_{\gamma}$, produce the shell oscillations $\delta g $ investigated semiclassically in this article. The frequencies are
determined by the classical actions $S_{\gamma}$ for the orbits,
while the phase is determined by the so called Maslov index $\mu_{\gamma}$
\cite{BBBook}. 

Specifically for the isotropic HO in $D$ dimensions,
we can write the trace formula representing the HO spectrum on a complex form to be used later \cite{BBBook,Ogren}
\begin{equation}
g\left(E\right)=\frac{1}{\hbar\omega}\frac{1}{\left(D-1\right)!}\prod_{j=1}^{D-1}\left(\frac{E}{\hbar\omega}-\frac{D}{2}+j\right)\,\text{Re}\left\{ \sum_{k=-\infty}^{\infty}(-1)^{Dk}e^{2\pi ikE/\hbar\omega}\right\} ,\: E>0.\label{eq:HOTF}
\end{equation}
The above equation is identically equal to the HO spectrum when viewing it as a train of delta spikes, each centered at the positions
(\ref{eq:HOQME}), and normalized to the degeneracy factor (\ref{eq:HOQMd_n}).
Moreover, the prefactor in (\ref{eq:HOTF}), corresponding to the $k=0$ term,
is equal to the extended Thomas-Fermi DOS \cite{BBBook,Ogren}, i.e., the first term in (\ref{eq:GeneralTF}). Its leading term in $\hbar^{-1}$ is the Thomas-Fermi DOS \cite{bermo}.
Moreover, the exponent of the summand in (\ref{eq:HOTF}) is in agreement
with the classical action of a primitive HO orbit being $S_{0}=2\pi E/\omega$,
and the Maslov index of the HO being zero \cite{BBBook}.

\section{Perturbation of the Harmonic oscillator}

In this article we consider perturbations to the HO of the form
\begin{equation}
\Delta H=\varepsilon\left|\mathbf{q}\right|^{2\alpha},\quad\alpha\in \mathbb{N},\quad\left|\mathbf{q}\right|^{2}=\sum_{j=1}^{D}q_{j}^{2},\label{eq:MonomialPerturbation}
\end{equation}
where the small parameter $\varepsilon$ has the dimension of $E/R_{0}^{2\alpha}$.
From \eqref{eq: HarmonicHamiltonian} and \eqref{eq:MonomialPerturbation}
we obtain the full Hamiltonian under study here
\begin{equation}
H\left(\mathbf{q},\mathbf{p}\right)=H_{0}+\Delta H=\frac{1}{2}\left(\left|\mathbf{p}\right|^{2}+\omega^{2}\left|\mathbf{q}\right|^{2}\right)+\varepsilon\left|\mathbf{q}\right|^{2\alpha}.\label{eq:FullHamiltonian}
\end{equation}
Note that for the special case of quartic perturbation, $\alpha=2$, the
Hamiltonian (\ref{eq:FullHamiltonian}) have been studied in two spatial
dimensions ($D=2$) in \cite{crpert}, and for $D=3$ in \cite{JoPA2005}.
In the present article we treat in detail an arbitrary even monomial
perturbation in $\left|\mathbf{q}\right|$, in any dimension $D$,
and in addition give an example of a realistic polynomial perturbation
for $D=3$.
The space of symmetries for this Hamiltonian is the set of orthogonal matrices $\mathrm{O}(D)$, a smaller space than $\mathrm{U}(D)$ for the unperturbed case. If we identify a solution with itself traversed backwards, the symmetries reduces to $\mathrm{SO}(D)$, the set of orthogonal matrices of determinant positive one. The resulting space $\mathrm{SO}(D)/\mathrm{SO}(2)$ is not so simple to describe,
and the lack of an explicit solution to (\ref{eq:FullHamiltonian}) makes it impossible to completely describe the manifold of the constant energy solutions in the general case.

\subsection{Perturbative Trace Formula} \label{sec: PertubativeTraceFormula}

We now turn our focus to the quantum mechanical energy spectrum of
the perturbed HO, with the goal to obtain a semiclassical trace formula
for the DOS of the Hamiltonian \eqref{eq:FullHamiltonian} within
first order perturbation theory. Starting from the HO trace formula
\eqref{eq:HOTF} and including a complex modulation factor $\mathcal{M}_k$ in the sum,
we define the perturbative trace formula to \eqref{eq:FullHamiltonian}
according to \cite{crpert}
\begin{equation}
g_{\mathrm{pert}}(E)\equiv\left(\hbar\omega\right){}^{-D}\frac{E^{D-1}}{\left(D-1\right)!}\,\text{Re}\left\{ \sum_{k=-\infty}^{\infty}(-1)^{Dk}\mathcal{M}_{k}e^{2\pi ikE/\hbar\omega}\right\} ,\: E>0. \label{eq:PerturbativeTraceFormula}
\end{equation}
The prefactor above only contains the leading order term in $\hbar^{-1}$ of
the extended Thomas-Fermi DOS for the unperturbed HO, see the prefactor
in \eqref{eq:HOTF}, in accordance with the order of the perturbative theory in use. 
The modulation factor in \eqref{eq:PerturbativeTraceFormula}
is generally defined according to \cite{BBBook,crpert}
\begin{equation}
\mathcal{M}_{k}\left(E,\varepsilon,D,\alpha,\omega \right)=\langle e^{ik\Delta S_\gamma/\hbar}\rangle_{\gamma\in\mathbb{C}P^{D-1}},\label{eq:DefModFactor}
\end{equation}
where $\gamma$ ranges over all classical periodic 
orbit of energy $E$ for the unperturbed HO, the last four variables are system dependent parameters.  $\Delta S$ is the lowest order term, with respect to $\varepsilon$, of the action in the perturbed system, see section \ref{sec: PerturbativeAction}.
In the following we shall calculate this expression explicitly. Using the Hopf map briefly described in section \ref{sec: ClassicalHO}, $\Delta S_\gamma$ induces a map on $S^{2D-1}$, given by $(\Delta S_\gamma \circ \pi_{\mathrm{Hopf}})$. Notationally we shall not distinguish between the two. Notice that $\Delta S_\gamma$ is constant on the fiber $S^1$. Rewriting,
\begin{align}
\mathcal{M}_{k}&=\frac{1}{\text{Vol}(\mathbb{C}P^{D-1})}\int_{\mathbb{C}P^{D-1}}e^{ik\Delta S_\gamma/\hbar}d\mathrm{vol}_{\mathrm{FS}}\nonumber \\
&=\frac{1}{ \text{Vol}(S^1) \text{Vol}(\mathbb{C}P^{D-1})}\int_{S^{2D-1}}e^{ik\Delta S_\gamma/\hbar}d\mathrm{vol}_{\mathrm{can}}, \label{eq: ModulationFactorSphereIntegral}
\end{align}
where $d\mathrm{vol}_{\mathrm{FS}}$ is the Fubini-Study volume form, $d\mathrm{vol}_{\mathrm{can}}$ is the volume form of the canonical round metric on $S^{2D-1}$, $\text{Vol}(S^1)=2\pi$, and $ \text{Vol}(\mathbb{C}P^{D-1}) = \pi^{D-1}/ \left( D-1 \right)!$. The spherical integral is significantly easier to compute, see section \ref{sec: ReductionModulationFactor}.

\subsection{Generalised angular momentum}
To analyse the perturbed system, we will use conserved quantities. Since $\mathrm{SO}(D)$ is a Lie group, i.e., a continuous group of symmetries, one can utilize Noether's theorem to directly compute conserved quantities. To this end we define a generalised angular momentum operator $\mathbf{L}:\mathbb{R}^{D}\times\mathbb{R}^{D}\rightarrow\mathbb{R}^{\frac{D(D-1)}{2}}$ according to
\begin{equation}
\mathbf{L}\left(\mathbf{q},\mathbf{p}\right)=(\dots,p_{j}q_{k}-p_{k}q_{j},\dots),\quad j,k=1...D,\; j\neq k.\label{eq:DefGenAngMom}
\end{equation}
That is, all combinations of the coordinates from $\mathbf{q}$ and
$\mathbf{p}$. Using Noether's theorem as given in \cite{GeometricMechanics}, one can show that all the coordinates are conserved for systems with $\mathrm{SO}(D)$ symmetry. Hence, this is just a generalization of the well known situation where angular momentum is preserved in three dimensional systems ($D=3$) with the rotational symmetry expressed by $\mathrm{SO}(3)$ invariance. Explicitly calculating the length of $\mathbf{L}$ from (\ref{eq:DefGenAngMom})
reveals that the following identity 
\begin{equation}
|\mathbf{L}|^{2}=\left|\mathbf{q}\right|^{2}\left|\mathbf{p}\right|^{2}-\left(\mathbf{q}\cdot\mathbf{p}\right)^{2}=\left|\mathbf{q}\right|^{2}\left|\mathbf{p}\right|^{2}\left(1-\cos^{2}\theta\right)=\sin^{2}\theta\left|\mathbf{q}\right|^{2}\left|\mathbf{p}\right|^{2},
\end{equation}
generally holds, just as in the common case where $\mathbf{q}, \mathbf{p} \in   \mathbb{R}^3$. Hence, we can define the conserved total angular momentum in $D$ dimensions as the area spanned by $\mathbf{q}, \mathbf{p}  \in  \mathbb{R}^D$:
\begin{equation}
L\equiv|\mathbf{L}|=\sin\theta\left|\mathbf{q}\right|\left|\mathbf{p}\right|, \label{def_of_L_in_terms_of_theta}
\end{equation}
where $\theta$ is the angle between the two vectors $\mathbf{q}$ and $\mathbf{p}$. 

\subsection{The perturbative action}
\label{sec: PerturbativeAction}
We here concentrate on the perturbative classical action $\Delta S_\gamma$, that occurs in the
exponent of (\ref{eq:DefModFactor}). 

In order to obtain a scaling for the perturbative action, we consider the following expansion 
\begin{align}
S &= \oint_\gamma  \mathbf{p} d \mathbf{q}  \sim 4 \int_0^{R_0} \sqrt{ 2E - \omega^2 r^2 - 2 \varepsilon r^{2 \alpha} } dr \simeq S_0 + \Delta S + \mathcal{O} \left( \varepsilon^2 \right)  \nonumber \\
&= \frac{2 \pi E}{\omega} - \varepsilon \frac{2^{\alpha+1} \sqrt{ \pi } \Gamma \left( \alpha + \frac{1}{2} \right) E^\alpha }{ \Gamma \left( \alpha + 1 \right) \omega^{2 \alpha +1 }}+ \mathcal{O} \left( \varepsilon^2 \right),
\end{align}
i.e., with the curve $\gamma$ corresponding to a classical diameter orbit. 
From the above result, we define the following scale of the first order perturbative action $\Delta S$ to be used later
\begin{equation}
\sigma_\alpha \equiv \varepsilon \frac{ 2 \pi E^\alpha}{ \omega^{2 \alpha + 1}} = \varepsilon \frac{  \pi R_0^{2\alpha}}{ 2^{\alpha-1}\omega},
\label{Scaling_for_Delta_S}
\end{equation}
such that $\sigma_\alpha / \hbar$ is dimensionless.

According to the first order semiclassical perturbation theory given in \cite{crpert},
we generally have

\begin{equation}
\Delta S_\gamma=-\oint_{\gamma}\Delta Hdt=-\varepsilon\int_{0}^{\frac{2\pi}{\omega}}\left|\mathbf{q}(t)\right|^{2\alpha}dt.\label{eq:FirstOrderAction}
\end{equation}
In earlier work, were more specific perturbations have been treated,
the calculation of \eqref{eq:FirstOrderAction} have been performed
with brute force methods. Involving for example specific parametrisations
of the periodic orbits on a hyper-sphere or a complex projective space
\cite{JoPA2005}, depending on the dimension $D$. The intention here is to avoid these technical calculations and use a more geometrical approach, independent of $\alpha$ and $D$. Ending with a reduced version of the trace formula in \eqref{eq:PerturbativeTraceFormula}, with an explicit dependence on the parameter space.

\begin{figure}
\centering{}

   \def\svgwidth{0.5\textwidth}
   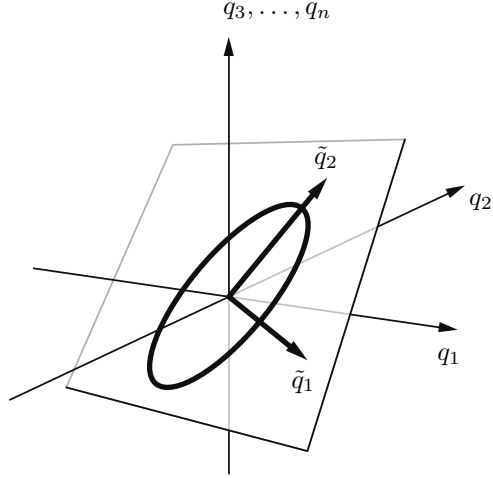

\caption{Illustration of a coordinate
system for which the orbits in configuration space can be written
on the form (\ref{eq: CanonicalOrbit}). \label{figure1} }
\end{figure}

To compute the circulation integral \eqref{eq:FirstOrderAction} for
classical periodic orbits $\mathbf{q}\left(t\right)$ {[}such
as (\ref{eq:HamiltonsEquationAndSolution}) or (\ref{eq:AlternativeParametrisation}){]}, consider a change of coordinates to a canonical form. As discussed in section \ref{sec: ClassicalHO}, the orbits are ellipses in the configuration
space. Hence for any orbit there exists an orthogonal change of coordinates,
such that $\mathbf{q}\left(t\right)$ can be written 
\begin{equation}
\mathbf{\tilde{q}}\left(t\right)=\left[a\cos(\omega t),b\sin(\omega t),0,...,0\right],\label{eq: CanonicalOrbit}
\end{equation}
for some constants $a,b\in\mathbb{R}$, see figure \ref{figure1}.

The energy and the total angular momentum are still conserved for the perturbed Hamiltonian
\eqref{eq:FullHamiltonian}.
Using (\ref{eq: CanonicalOrbit}) they are easily found to be 
\begin{equation}
\frac{2E}{\omega^2} = R_{0}^{2}=a^{2}+b^{2},\quad \frac{L^{2}}{\omega^2}=a^{2}b^{2}.
\label{eq: ConservedQuantitiesCanonicalOrbit}
\end{equation}
Solving for $a$ and $b$ in terms of the conserved quantities
$R_{0}$ and $L$ yields 
\begin{equation}
a^{2}=\frac{R_{0}^{2}}{2}+\sqrt{\frac{R_{0}^{4}}{4}-\frac{L^{2}}{\omega^{2}}},\quad b^{2}=\frac{R_{0}^{2}}{2}-\sqrt{\frac{R_{0}^{4}}{4}-\frac{L^{2}}{\omega^{2}}}.\label{eq: EllipticalAxisInRL}
\end{equation}
The action integral (\ref{eq:FirstOrderAction}) calculated in the $\mathbf{\tilde{q}}$-coordinates becomes
\begin{equation}
\Delta S=-\varepsilon\int_{0}^{\frac{2\pi}{\omega}}\left[a^{2}\cos^{2}(\omega t)+b^{2}\sin^{2}(\omega t)\right]^{\alpha}dt.
\end{equation}
We substitute $s=\omega t$ and apply the Binomial theorem
\begin{equation}
\Delta S=-\frac{\varepsilon}{\omega}\int_{0}^{2\pi}\sum_{k=0}^{\alpha}\binom{\alpha}{k}a^{2k}\cos^{2k}\left(s\right)\; b^{2\alpha-2k}\sin^{2\alpha-2k}\left(s\right)ds.\label{eq:Apply_the_Binomial_theorem}
\end{equation}
By a direct calculation using the formulas for trigonometric integrals in \cite{AbramowitzStegun1972}, we obtain the following identity for $0\leq k\leq\alpha$
\begin{equation}
\frac{1}{2\pi}\int_{0}^{2\pi}\sin^{2\alpha-2k}\left(s\right)\cos^{2k}\left(s\right)ds=\frac{(2k-1)!!\left[2\alpha-\left(2k+1\right)\right]!!}{\left(2\alpha\right)!!},
\end{equation}
where $n!!$ is the double factorial of $n$ (not to be confused with twice factorial $(n!)!$). From (\ref{eq:Apply_the_Binomial_theorem}) above 
\begin{equation}
\Delta S=-\frac{2\pi\varepsilon}{\omega}\sum_{k=0}^{\alpha}\frac{\alpha!(2k-1)!!\left[2\alpha-\left(2k+1\right)\right]!!}{k!(\alpha-k)!\left(2\alpha\right)!!}a^{2k}b^{2\alpha-2k}.\label{eq:DeltaSDoubleFactorial}
\end{equation}
 For a more convenient notation, we define coefficients $I_{\alpha}^{k}$
in (\ref{eq:DeltaSDoubleFactorial}), such that
\begin{equation}
\Delta S=-\frac{2\pi\varepsilon}{\omega}\sum_{k=0}^{\alpha}I_{\alpha}^{k}a^{2k}b^{2\alpha-2k},\label{eq: DeltaSAlphaEven}
\end{equation}
where we note that $I_{\alpha}^{k}=I_{\alpha}^{\alpha-k}$. Due to
this symmetry we can reduce the expression \eqref{eq:DeltaSDoubleFactorial},
depending on whether $\alpha$ is even or odd. As we are interested in integrating this expression over the orbits of the HO, we would like to use \eqref{eq: ConservedQuantitiesCanonicalOrbit} to rewrite this into an expression in $R_0$ and $L$, since these are easily obtained given a specific orbit. To this end let $\lfloor x\rfloor$
denotes the floor of $x$, i.e., the largest integer fulfilling $\lfloor x\rfloor\leq x$,
and we can then write
\begin{equation}
\Delta S=\left\{ \begin{array}{l}
-\frac{2\pi\varepsilon}{\omega}\sum_{k=0}^{\lfloor\alpha/2\rfloor}I_{\alpha}^{k}\left(a^{2k}b^{2\alpha-2k}+a^{2\alpha-2k}b^{2k}\right)\quad,\alpha\text{ odd},\\
-\frac{2\pi\varepsilon}{\omega}\sum_{k=0}^{\alpha/2-1}I_{\alpha}^{k}\left(a^{2k}b^{2\alpha-2k}+a^{2\alpha-2k}b^{2k}\right)+I_{\alpha}^{\alpha/2}a^{\alpha}b^{\alpha}\quad,\alpha\text{ even}.
\end{array}\right.\label{eq:DeltaS_odd_and_even}
\end{equation}
Here combinations of $a^{2}$ and $b^{2}$ can be replaced by the
expressions in \eqref{eq: EllipticalAxisInRL}, such that
\begin{equation}
a^{2k}b^{2\alpha-2k} + a^{2\alpha-2k}b^{2k} =  \frac{R_0^{2\alpha}}{2^{\alpha}}\sum_{l=0}^{k}\sum_{p=0}^{\alpha-k} K_{l,p}^{\alpha,k}  \left(1-\frac{4L^{2}}{\omega^2 R_0^4}\right)^{\frac{l+p}{2}}, \label{eq:ExpressionIn_a_and_b}
\end{equation}
with the constants
\begin{equation}
K_{l,p}^{\alpha,k} = \binom{k}{l}\binom{\alpha-k}{p} \left[ \left(-1\right)^l + \left(-1\right)^p \right].  \label{eq:K_lp_ak}
\end{equation}
The expression \eqref{eq:ExpressionIn_a_and_b} is in fact a polynomial
in the two constants of the motion, $R_{0}^{2}$ and $L^{2}$. Observe that $K^{\alpha,k}_{l,p} = 0$ if the parity of $l$ and $p$ is not the same. Hence only terms in the double sum with $l+p$ even will be non-zero. The last term in \eqref{eq:DeltaS_odd_and_even}, for $\alpha$ even is 
\begin{equation}
a^\alpha b^\alpha = \frac{L^{\alpha}}{\omega^\alpha}.
\label{eq:ExpressionIn_ab}
\end{equation}
Defining a dimensionless angular momentum according to
\begin{equation}
\tilde{L} \equiv \frac{2 L}{ \omega R_0^2},
\label{eq:definition_of_L_tilde}
\end{equation} 
and inserting (\ref{eq:ExpressionIn_a_and_b}) and \eqref{eq:ExpressionIn_ab}
into \eqref{eq:DeltaS_odd_and_even}, we can transform \eqref{eq: DeltaSAlphaEven} into the form
\begin{equation}
\Delta S= - \sigma_\alpha \sum_{j=0}^{\lfloor\alpha/2\rfloor}a_{j} \tilde{L}^{2j},\label{eq:DeltaS_as_a_function_of_L}
\end{equation}
for some coefficients $a_{j}(\alpha)$, which only depends on the order of the perturbation $\alpha$, but not on the spatial
dimension $D$ of the system, see table \ref{table_for_Delta_S} for examples. 
First, we can see that for
$\alpha=1$, we have $\lfloor\alpha/2\rfloor=0$, such that $i k \Delta S=0$
and then (\ref{eq:DefModFactor}) gives $\mathcal{M}_{k}\equiv1$ in any dimension $D$. 
Hence, the perturbative trace formula
(\ref{eq:PerturbativeTraceFormula}) for the oscillating part of the DOS naturally give no information about the
frequency shift $\omega_{\textnormal{{eff}}} \equiv \sqrt{\omega^{2}+2\varepsilon}$
of a harmonic perturbation \cite{OgrenPRA2007}. 
The shift of the main HO levels can be taken into account by a (perturbative) calculation of the smooth TF DOS of the system as outlined in appendix C of \cite{JoPA2005}.
Clearly $\mathcal{M}_{k}\equiv1$
also for $\varepsilon=0$ by definition, and the corresponding
perturbed trace formula uniformly restores the unperturbed TF
in the limit $| \varepsilon | \rightarrow0$. 
As a non-trivial example, take $\alpha = 2$, i.e., a quartic perturbation, to obtain
\begin{equation}
\Delta S =-\varepsilon\frac{\pi R_{0}^{4}}{4\omega}\left(3-\frac{4L^{2}}{\omega^{2}R_{0}^{4}}\right) = - \sigma_2 \frac{1}{2} \left( 3 -  \tilde{L}^2 \right),
\label{Delta_S_for_alpha_2}
\end{equation} 
with $\sigma_2$ from (\ref{Scaling_for_Delta_S}) and $\tilde{L}$ from (\ref{eq:definition_of_L_tilde}).
This is in agreement with what have implicitly been derived by Brack \textit{et.
al.} in three-dimensions \cite{JoPA2005} and by Craigh in two-dimensions \cite{crpert}.
However, the approach presented here has no limitations for $\alpha$ in any dimension, such that for example $\alpha = 3$ gives
\begin{equation}
\Delta S = -\varepsilon \frac{\pi R_0^6}{8 \omega} \left( 5 - \frac{12 L^2}{\omega^2 R_0^4} \right) = - \sigma_3 \frac{1}{2} \left( 5 - 3 \tilde{L}^2 \right). \label{Delta_S_for_alpha_3}
\end{equation}
We summarize the rest of the first ten cases in table \ref{table_for_Delta_S}.
A pattern seem to emerge, and in the general case we conjecture that the following identity holds
\begin{equation}
\Delta S = - \sigma_\alpha \tilde{L}^\alpha  P_\alpha \left( \frac{1}{\tilde{L}} \right), \label{Delta_S_conjecture}
\end{equation}
where $P_\alpha$ denote the Legendre polynomial of order $\alpha$. This provides an explicit form of the coefficients $a_j$ in \eqref{eq:DeltaS_as_a_function_of_L}. We have not proven (\ref{Delta_S_conjecture}) but confirmed that it holds for $\alpha \leq 1000$ with a CAS software. In section \ref{sec: SPA} the zeros of $\Delta S(\tilde{L})$ will be important, and we can then utilize that the zeros of $P_\alpha$ are well understood.
\begin{table}[ht]
\caption{Results for the perturbative action $ -\Delta S/\sigma_\alpha =\sum_{j=0}^{\lfloor\alpha/2\rfloor}a_{j} \tilde{L}^{2j}$, for a monomial potential $ \varepsilon r^{2 \alpha} $ in arbitrary dimensions.}
\begin{tabular}{ || c | c ||} 
\hline \hline
 $\alpha = 4$ & $\alpha =  5$  \\ 
\hline
 $ \frac{1}{8} \left( 35 - 30 \tilde{L}^2 + 3 \tilde{L}^4 \right) $ & $ \frac{1}{8} \left( 63 - 70 \tilde{L}^2 + 15 \tilde{L}^4 \right) $  \\ 
\hline 
\end{tabular}
\begin{tabular}{ || c | c ||} 
\hline 
 $\alpha =  6$ & $\alpha =  7$  \\ 
\hline
 $ \frac{1}{16} \left( 231 - 315 \tilde{L}^2 + 105 \tilde{L}^4 - 5 \tilde{L}^6 \right) $ & $ \frac{1}{16} \left( 429 - 693 \tilde{L}^2 + 315 \tilde{L}^4 - 35 \tilde{L}^6 \right) $\\ 
\hline 
\end{tabular}
\begin{tabular}{ || c ||} 
\hline 
$\alpha =  8$  \\ 
\hline
$  \frac{1}{128} \left( 6435 - 12012 \tilde{L}^2 + 6930 \tilde{L}^4 - 1260 \tilde{L}^6 + 35 \tilde{L}^8 \right) $ \\ 
\hline 
\end{tabular}  \\
\begin{tabular}{ || c ||} 
\hline 
 $\alpha =  9$  \\ 
\hline
$  \frac{1}{128} \left( 12155 - 25740 \tilde{L}^2 + 18018 \tilde{L}^4 - 4620 \tilde{L}^6 + 315 \tilde{L}^8 \right) $ \\ 
\hline
\end{tabular} \\
\begin{tabular}{ || c ||} 
\hline 
 $\alpha =  10$  \\ 
\hline
$  \frac{1}{256} \left( 46189 - 109395 \tilde{L}^2 + 90090 \tilde{L}^4 - 30030 \tilde{L}^6 + 3465 \tilde{L}^8 - 63 \tilde{L}^{10} \right) $ \\ 
\hline \hline
\end{tabular}
\label{table_for_Delta_S}
\end{table}

\subsection{Reduction of the modulation factor}\label{sec: ModulationFactor}
\label{sec: ReductionModulationFactor}

Recall that we are holding $E$ fixed (hence also $R_0$), so only $L$ changes in \eqref{eq:DeltaS_as_a_function_of_L} as $\gamma$ varies in the family of fixed energy periodic orbits. 
In particular a so called diameter orbit have $L=0$, while
the maximum of $L$ is obtained for a circular orbit, where $a^2 = b^2 = R_0^2/2$, corresponding to zero radial momentum.

Now focusing on calculating the orbit invariant $L(\mathbf{q_0},\mathbf{p_0})$, for the variables $(\mathbf{q_0}/R_0,\mathbf{p_0}/ \omega R_0) \in S^{2D-1}$, used in the spherical integral for the modulation factor in \eqref{eq: ModulationFactorSphereIntegral}.
It will be beneficial to consider the sphere $S^{2D-1}$ as the following set
\begin{equation}
S^{2D-1}=\left\{ \left[ \cos\left(\frac{\varphi}{2}\right)\,\mathbf{e}_{\mathbf{q}},\sin\left(\frac{\varphi}{2}\right)\,\mathbf{e}_{\mathbf{p}} \right] \;|\;\mathbf{e}_{\mathbf{q}},\mathbf{e}_{\mathbf{p}}\in S^{D-1},\varphi\in[0,\pi] \right\}.
\label{eq: SphereSplitting}
\end{equation}
Using (\ref{eq: SphereSplitting}), we can rewrite (\ref{def_of_L_in_terms_of_theta}) according to
\begin{align}
L\left(\mathbf{q},\mathbf{p}\right)&=\omega R_{0}^{2}\cos\left(\frac{\varphi}{2}\right)\sin\left(\frac{\varphi}{2}\right)L\left(\mathbf{e}_{\mathbf{q}},\mathbf{e}_{\mathbf{p}}\right) \nonumber \\
&= \frac{\omega R_{0}^{2}}{2}\sin\left(\varphi\right)L\left(\mathbf{e_{q}},\mathbf{e_{p}}\right)=\frac{\omega R_{0}^{2}}{2}\sin\left(\varphi\right)\sin\left(\theta\right),  \label{eq: AngulatorMomentumPhiTheta}
\end{align}
where $\theta$ is the angle between the two vectors $\mathbf{e}_{\mathbf{q}}$
and $\mathbf{e}_{\mathbf{p}}$ of unit length.
\begin{figure}
\centering{}
  \def\svgwidth{0.45\textwidth}
  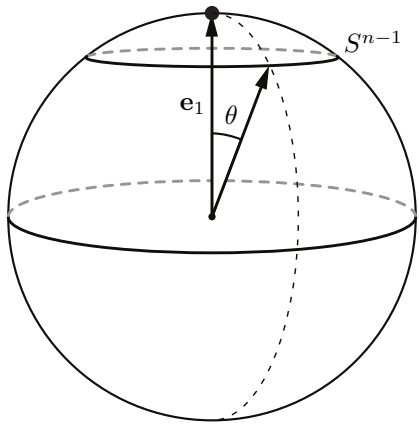

\caption{ Construction of $S^n$ as a warped product of $[0,\pi]$; a meridian, and $S^{n-1}$ of radius $\sin(\theta)$; a parallel. 
\label{figure2} }
\end{figure} 
That is, taking a $ \Delta S$ polynomial in $\tilde{L}$ from (\ref{Delta_S_for_alpha_2}), (\ref{Delta_S_for_alpha_3}), (\ref{Delta_S_conjecture}), or table \ref{table_for_Delta_S}, one should interchange $\tilde{L} \rightarrow \sin\left(\varphi\right)\sin\left(\theta\right) $ to obtain the form to be used in this section.

Given an energy $E$, the integrand of $\mathcal{M}_k$ in \eqref{eq: ModulationFactorSphereIntegral}, is now only dependent
on $L\left(\varphi,\theta\right)=|\mathbf{L}|$. The integration measure for the splitting of the sphere in \eqref{eq: SphereSplitting}, is given by
\begin{equation}
d\mathrm{vol}_{S^{2D-1}}=\frac{1}{ 2^{D}}\sin^{D-1}\left( \varphi \right)\; d\mathrm{vol}_{\mathbf{e}_{q}}d\mathrm{vol}_{\mathbf{e}_{p}}d\varphi.
\end{equation}
To integrate over the two smaller spheres, we use the observation that in the integrand, the only dependence of the variables is given by \eqref{eq: AngulatorMomentumPhiTheta}, and then only the angle between $\mathbf{e}_q$ and $\mathbf{e}_p$. With this in mind, consider $S^{n}$ as the set 
\begin{equation}
S^{n}=\left\{ \cos (\theta), \sin(\theta) \; \mathbf{e} \; |\; \mathbf{e} \in S^{n-1},\theta \in [0,\pi] \right\}, \label{Sn_as_the_set}
\end{equation}
see figure \ref{figure2} for an illustration. In the case of $S^2$ this reduces to the usual spherical coordinates. 
In the general case, the metric arising in this fashion is called a warped product structure of $S^n$, see \cite{ONeill}. Notice that $\theta $ exactly corresponds to the angle appearing in \eqref{eq: AngulatorMomentumPhiTheta}.
The integration measure induced by (\ref{Sn_as_the_set}) is
\begin{equation}
d \mathrm{vol}_{S^n} = \sin^{n-1}(\theta) d \mathrm{vol}_{S^{n-1}}.
\end{equation}
Collecting our results, with $n =D-1$, yields the following double integral in $\varphi$ and $\theta$
\begin{equation}
\mathcal{M}_{k}\left(E\right)=\frac{2 \left( D-1 \right)}{\pi}\int_{0}^{\pi/2}\int_{0}^{\pi/2}\sin^{D-1}\left( \varphi \right)\sin^{D-2}\left(\theta\right)\; e^{ik\Delta S\left(\varphi,\theta\right)/\hbar}d\varphi d\theta,  \label{eq:Modulationfactor}
\end{equation}
where 
\begin{equation}
\Delta S\left(\varphi,\theta\right) = - \sigma_\alpha \sum_{j=0}^{\lfloor\alpha/2\rfloor}a_{j} \sin^{2j}\left(\varphi\right)\sin^{2j}\left(\theta\right), 
\end{equation}
according to (\ref{eq:DeltaS_as_a_function_of_L}) and \eqref{eq: AngulatorMomentumPhiTheta}. 
The pre-factor in (\ref{eq:Modulationfactor}) was obtained by evaluating 
\begin{equation}
4 \frac{\mathrm{Vol}(S^{D-1}) \mathrm{Vol}(S^{D-2})}{2\pi \mathrm{Vol}(\mathbb{C}P^{D-1}) 2^D }= \frac{ 2 \left( D-1 \right) }{\pi}, 
\end{equation}
where the factor $4$ to the left above is due to the reduction of the two remaining upper integration limits that follows from the symmetry of the integrand.

In order to reduce (\ref{eq:Modulationfactor}) to a single integral we use new variables $\ell \in \left[0,1 \right]$ and $\vartheta \in \left[0,\pi/2 \right]$, defined according to
\begin{equation}
\ell = \sin(\varphi) \sin(\theta), \quad \cos(\varphi) = \sqrt{1-\ell^2} \: \sin(\vartheta).
\end{equation}
Simplifying the corresponding Jacobian to $\sin(\varphi) d\varphi d\theta = d\ell d\vartheta$, and using (\ref{Delta_S_conjecture}) for $\Delta S(\ell)$, we are left with
\begin{equation}
\mathcal{M}_k = (D-1) \int_0^1 \ell^{D-2} e^{-ik\sigma_\alpha \ell^{\alpha} P_\alpha ( \frac{1}{\ell} )/\hbar} d\ell. \label{OneDimensionalIntegral}
\end{equation}
Generally, exponentials of high orders, $\alpha \geq 4$, do not have known integrals. 
Restricting the discussion for a moment to the case where $a_j=0$ for $j \geq 2$, i.e., according to table \ref{table_for_Delta_S}, to perturbations with $\alpha=2,3$. One can show that (\ref{OneDimensionalIntegral}) can be expressed with help of a generalized hypergeometric function $_{p}F_{q}$ \cite{Bailey}
\begin{equation}
\mathcal{M}_k = e^{-i k (a_0+a_1) \sigma_\alpha / \hbar} \left[ 1 + \frac{2 z}{ D+1 }   + \frac{4 z^{2}  \: _{1}F_{ 1 }  \left( 1;\frac{ D+5}{2 } ; z \right) }{ \left( D+1 \right) \left( D+3 \right) }  \right], \label{eq:HyperModulationfactor}
\end{equation}
where the argument is $z=i k  \sigma_\alpha a_1 / \hbar$.

As specific examples we give in table \ref{table_for_M_k} the modulation factors valid for perturbations with $\alpha = 2,3$ for different dimensions $D=2,3, \hdots,7$. 
For odd dimensions $D$, the integral seem to always be expressible using elementary functions, and for even $D$ the error function (erf) can be used.

We note that the two-dimensional ($D=2$) case was in \cite{crpert} equivalently expressed in terms of Fresnel integrals.
\begin{table}[ht]
\caption{Modulation factors $\mathcal{M}_k$ according to (\ref{eq:HyperModulationfactor}), valid for the two monomial potentials $ \varepsilon r^{4} $ and $ \varepsilon r^{6} $ in different dimensions $D$.} 
\begin{tabular}{ || c | c ||} 
\hline \hline
 $D =2$ & $D =  3$   \\ 
\hline
 $ \frac{ \sqrt{\pi}}{2}     \frac{ \textnormal{erf} \left(  \sqrt{ \frac{i k \sigma_\alpha a_1}{\hbar} }  \right) } {\sqrt{ \frac{i k \sigma_\alpha a_1}{\hbar} }}  e^{-i k \sigma_\alpha a_0 / \hbar} $ & $ \frac{i \hbar}{k \sigma_\alpha a_1} \left(  e^{-i k \sigma_\alpha \left(  a_0 + a_1 \right) / \hbar} -  e^{-i k \sigma_\alpha a_0 / \hbar} \right) $  \\ 
\hline 
\end{tabular}
\begin{tabular}{ || c ||} 
\hline 
 $D =  4$   \\ 
\hline
 $   \frac{3 i \hbar}{4 k \sigma_\alpha a_1} \left( 2 e^{-i k \sigma_\alpha \left(  a_0 + a_1 \right) / \hbar} - \sqrt{\pi} \frac{ \textnormal{erf} \left(  \sqrt{ \frac{i k \sigma_\alpha a_1}{\hbar} }  \right) }{\sqrt{ \frac{i k \sigma_\alpha a_1}{\hbar} }} e^{- i k \sigma_\alpha a_0 / \hbar}  \right)
 $ \\ 
\hline 
\end{tabular}
\begin{tabular}{ || c ||} 
\hline 
 $D =  5$  \\ 
\hline
 $ \frac{2 \hbar}{k^2 \sigma_\alpha^2 a_1^2} \left( \left[ i k \sigma_\alpha a_1 + \hbar \right] e^{-i k \sigma_\alpha \left(  a_0 + a_1 \right) / \hbar}  - \hbar e^{-i k \sigma_\alpha a_0 / \hbar} \right)  $ \\ 
\hline
\end{tabular} \\
\begin{tabular}{ || c ||} 
\hline 
 $D =  6$  \\ 
\hline
 $  \frac{5 \hbar}{ 8 k^2 \sigma_\alpha^2 a_1^2}  \left( \left[ 4 i k \sigma_\alpha a_1 + 6 \hbar \right]  e^{-i k \sigma_\alpha \left(  a_0 + a_1 \right) / \hbar} - 3 \sqrt{\pi}  \hbar \frac{ \textnormal{erf} \left(  \sqrt{ \frac{i k \sigma_\alpha a_1}{\hbar} }  \right) }{\sqrt{ \frac{i k \sigma_\alpha a_1}{\hbar} }} e^{- i k \sigma_\alpha a_0 / \hbar}  \right)
 $ \\ 
\hline 
\end{tabular}\\
\begin{tabular}{ || c ||} 
\hline 
 $D =  7$  \\ 
\hline
 $ \frac{3 \hbar}{k^3 \sigma_\alpha^3 a_1^3} \left( \left[ i k^2 \sigma_\alpha^2 a_1^2  + 2 \hbar k \sigma_\alpha a_1    - 2 i  \hbar^2 \right]  e^{-i k \sigma_\alpha \left(  a_0 + a_1 \right) / \hbar}   + 2 i  \hbar^2  e^{-i k \sigma_\alpha a_0 / \hbar}  \right) $ \\ 
\hline \hline
\end{tabular} \\
\label{table_for_M_k}
\end{table}

\subsection{Stationary phase approximation}
\label{sec: SPA}
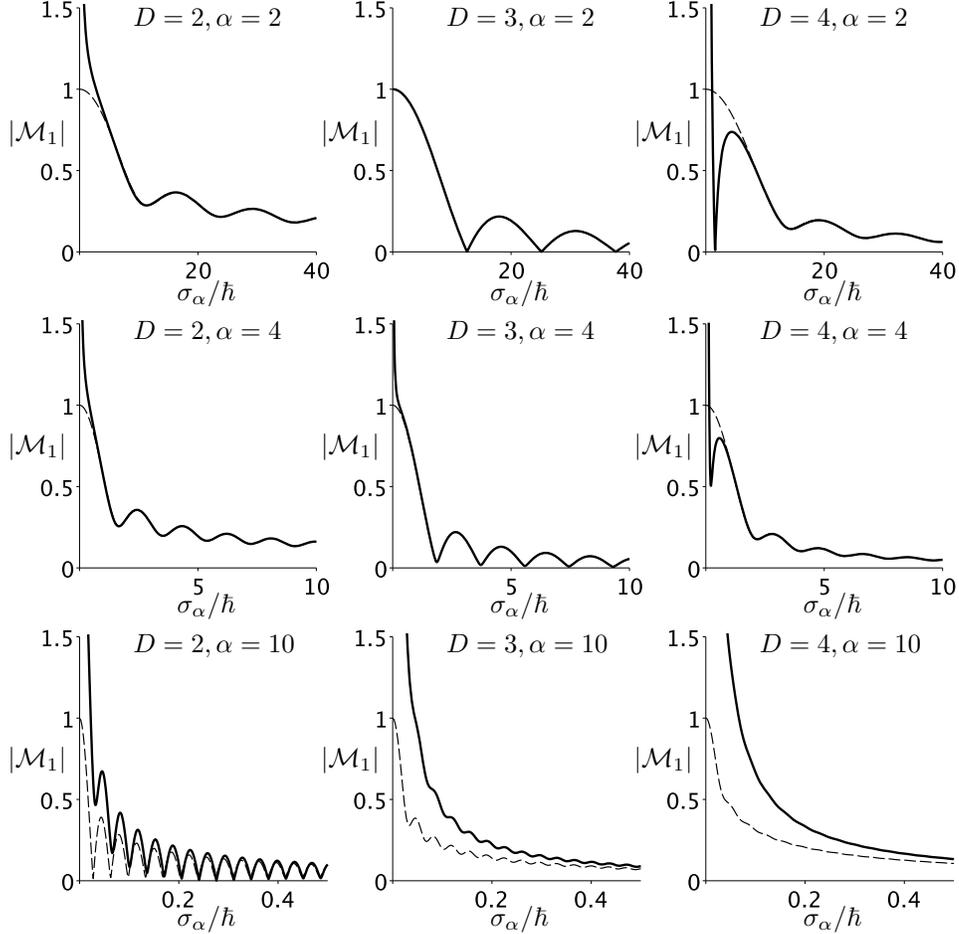
\begin{figure}
\centering{}
    \setlength{\unitlength}{118bp}%
  \begin{picture}(3,3)%
    	\put(-0.1,2){\includegraphics[width=1\unitlength]{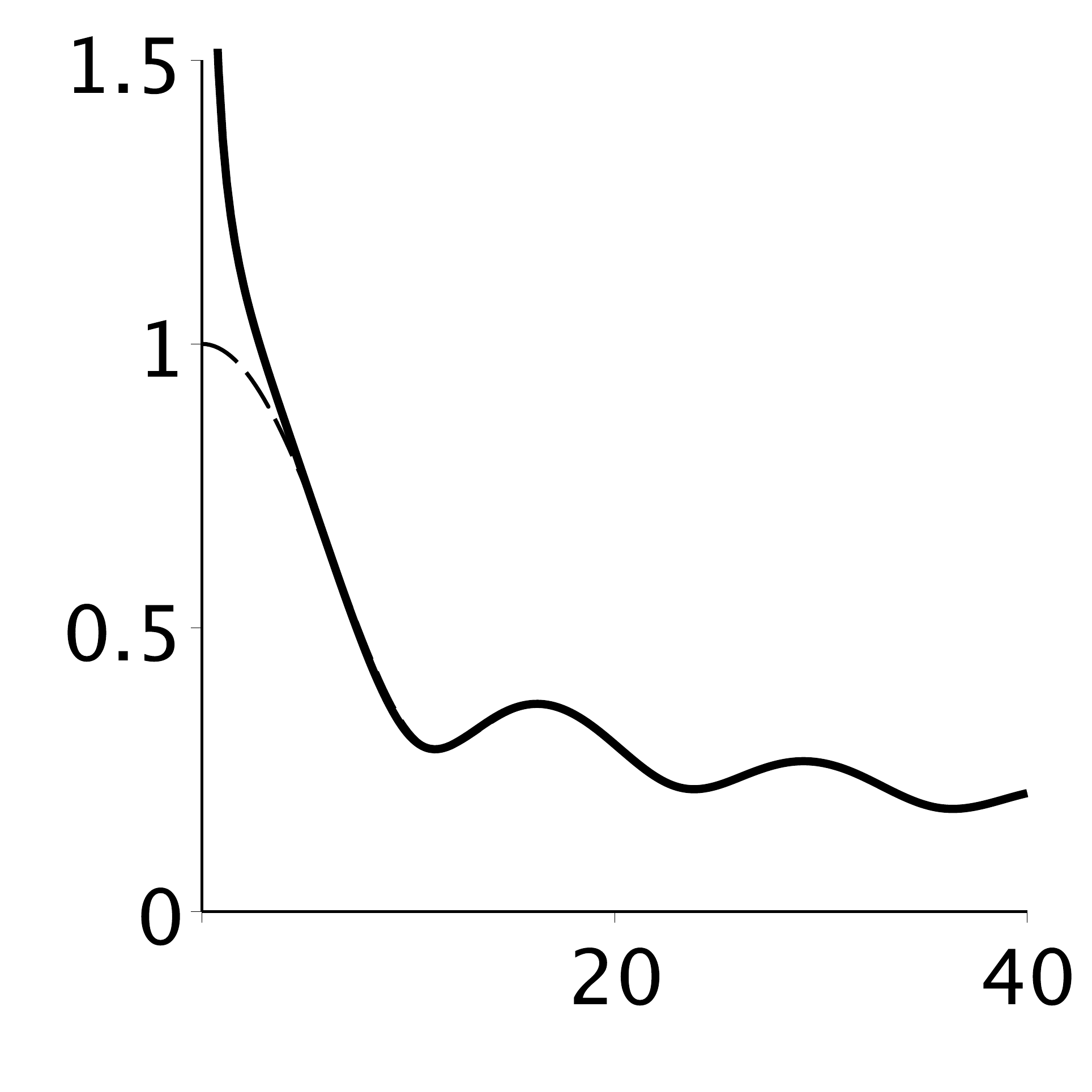}}%
    \put(0.2,2.87){\color[rgb]{0,0,0}\makebox(0,0)[lb]{ { $D=2,\alpha=2$} }}%
    \put(-0.2,2.5){\color[rgb]{0,0,0}\makebox(0,0)[lb]{ { $|\mathcal{M}_1|$} }}%
    \put(0.34,1.99){\color[rgb]{0,0,0}\makebox(0,0)[lb]{ { $\sigma_\alpha/\hbar$} }}%
    	\put(0.9,2){\includegraphics[width=1\unitlength]{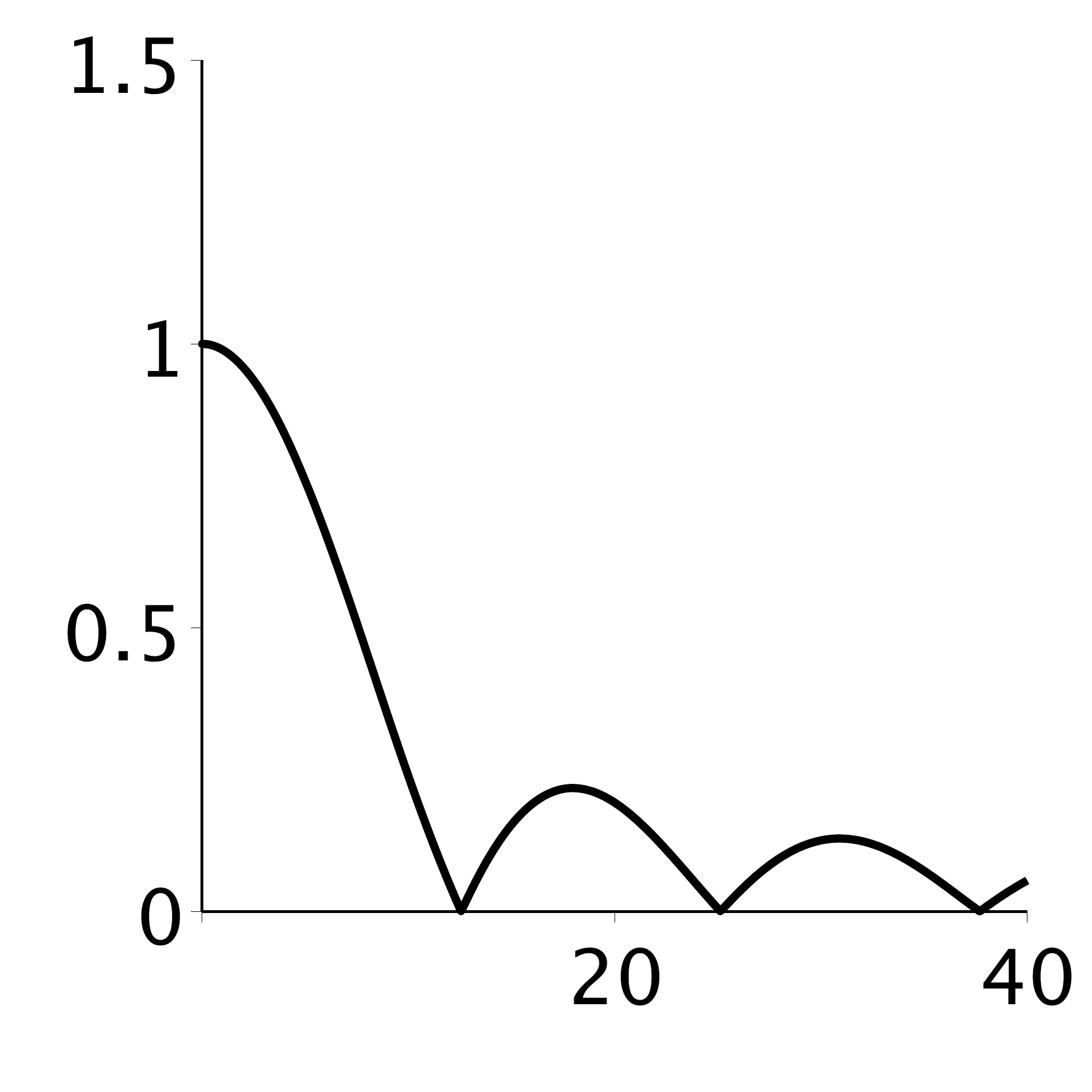}}%
    \put(1.2,2.87){\color[rgb]{0,0,0}\makebox(0,0)[lb]{ { $D=3,\alpha=2$} }}%
    \put(0.8,2.5){\color[rgb]{0,0,0}\makebox(0,0)[lb]{ { $|\mathcal{M}_1|$} }}%
    \put(1.34,1.99){\color[rgb]{0,0,0}\makebox(0,0)[lb]{ { $\sigma_\alpha/\hbar$} }}%        
    	\put(1.9,2){\includegraphics[width=1\unitlength]{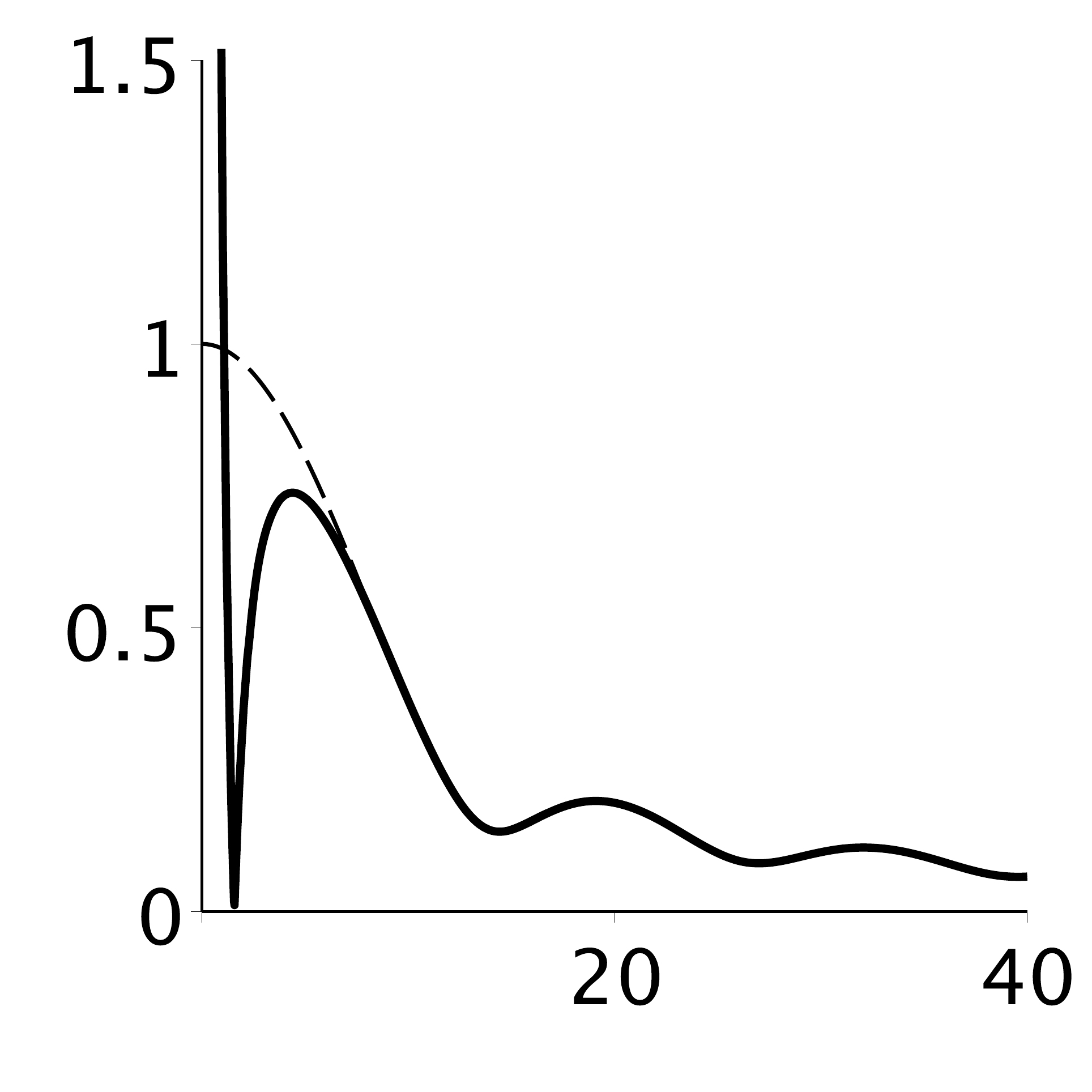}}%
    \put(2.2,2.87){\color[rgb]{0,0,0}\makebox(0,0)[lb]{ { $D=4,\alpha=2$} }}%
    \put(1.8,2.5){\color[rgb]{0,0,0}\makebox(0,0)[lb]{ { $|\mathcal{M}_1|$} }}% 
    \put(2.34,1.99){\color[rgb]{0,0,0}\makebox(0,0)[lb]{ { $\sigma_\alpha/\hbar$} }}%

    	\put(-0.1,0.99){\includegraphics[width=1\unitlength]{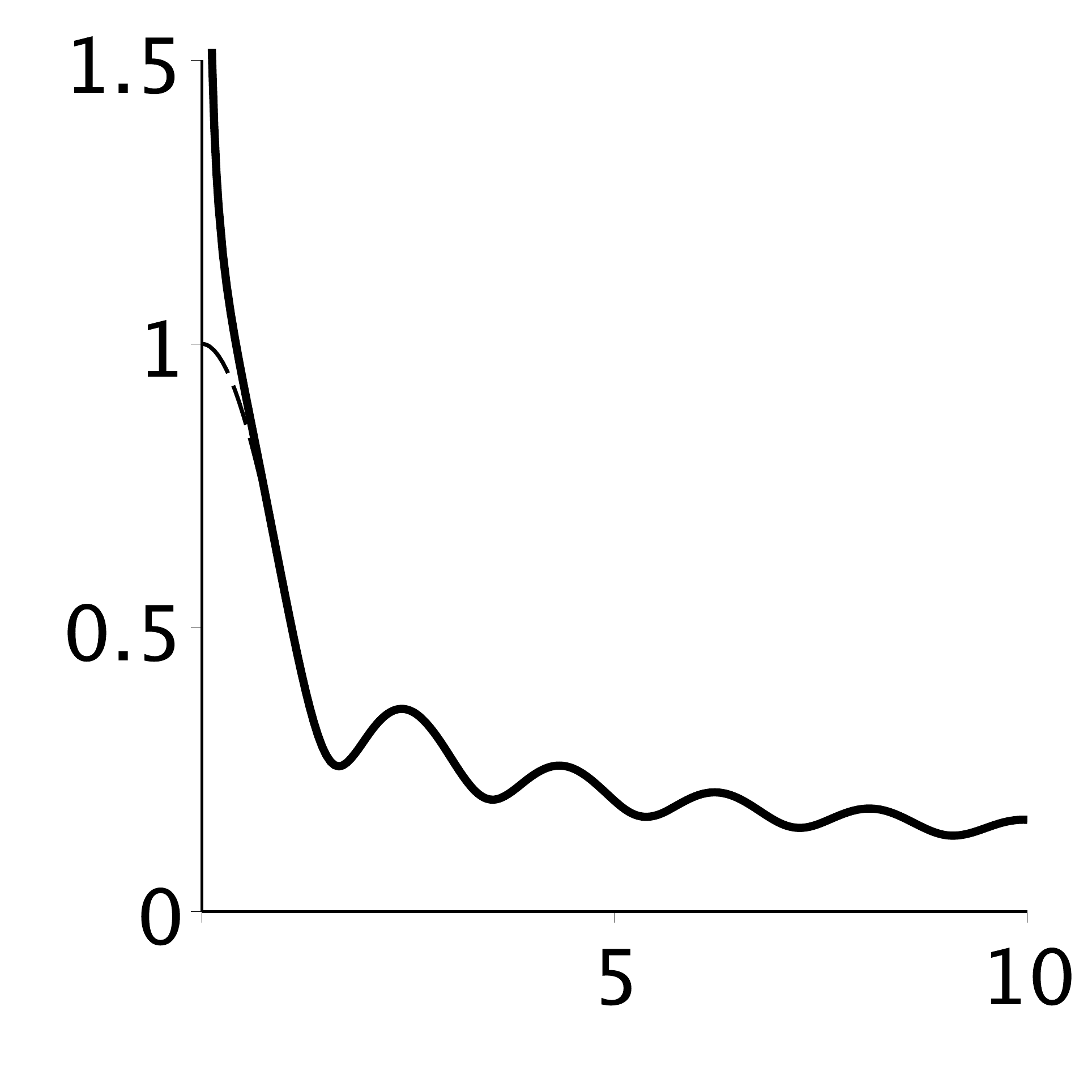}}%
    \put(0.2,1.87){\color[rgb]{0,0,0}\makebox(0,0)[lb]{ { $D=2,\alpha=4$} }}%
    \put(-0.2,1.5){\color[rgb]{0,0,0}\makebox(0,0)[lb]{ { $|\mathcal{M}_1|$} }}
    \put(0.34,0.99){\color[rgb]{0,0,0}\makebox(0,0)[lb]{ { $\sigma_\alpha/\hbar$} }}%
        \put(0.9,0.99){\includegraphics[width=1\unitlength]{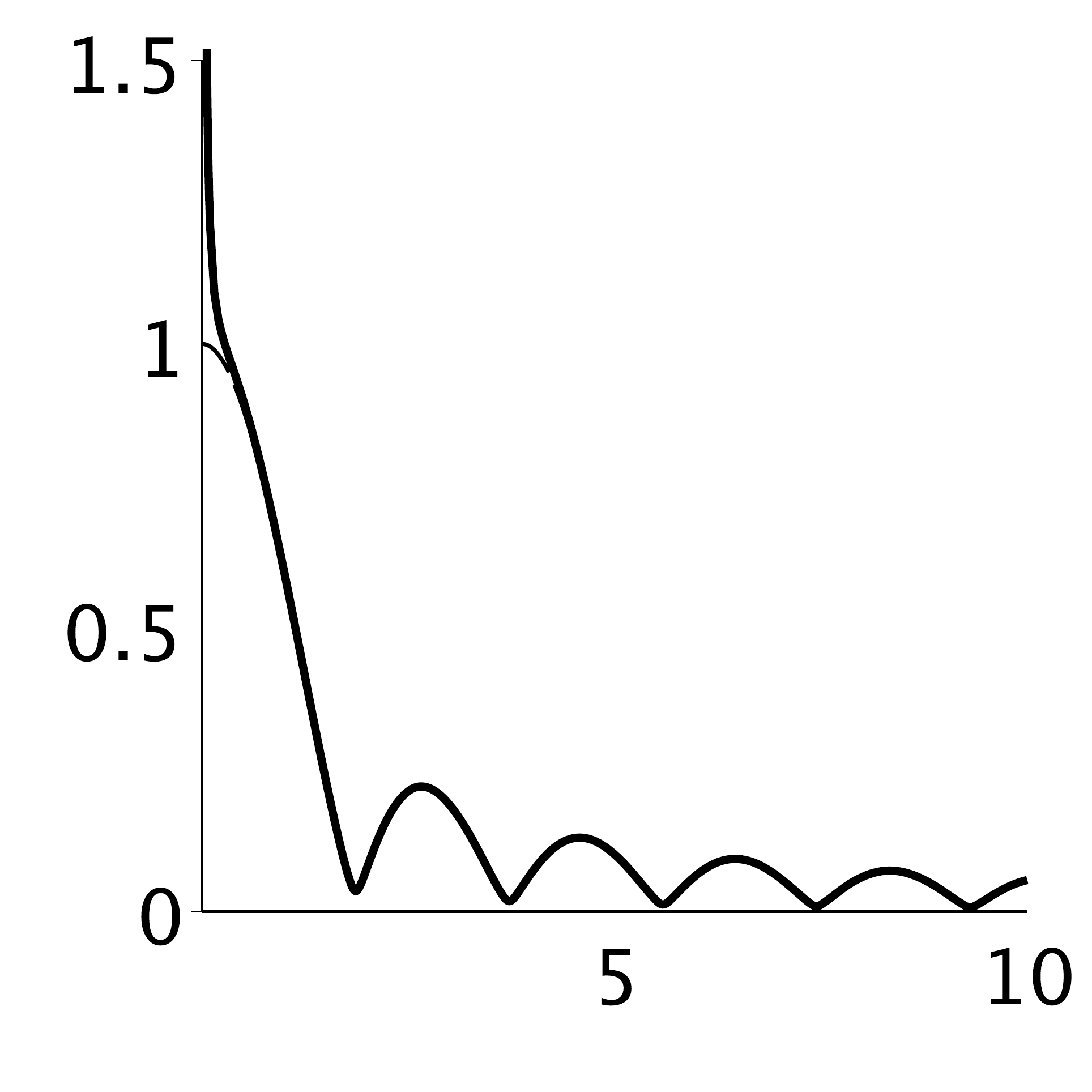}}%
    \put(1.2,1.87){\color[rgb]{0,0,0}\makebox(0,0)[lb]{ { $D=3,\alpha=4$} }}%
    \put(0.8,1.5){\color[rgb]{0,0,0}\makebox(0,0)[lb]{ { $|\mathcal{M}_1|$} }}
    \put(1.34,0.99){\color[rgb]{0,0,0}\makebox(0,0)[lb]{ { $\sigma_\alpha/\hbar$} }}%
        \put(1.9,0.99){\includegraphics[width=1\unitlength]{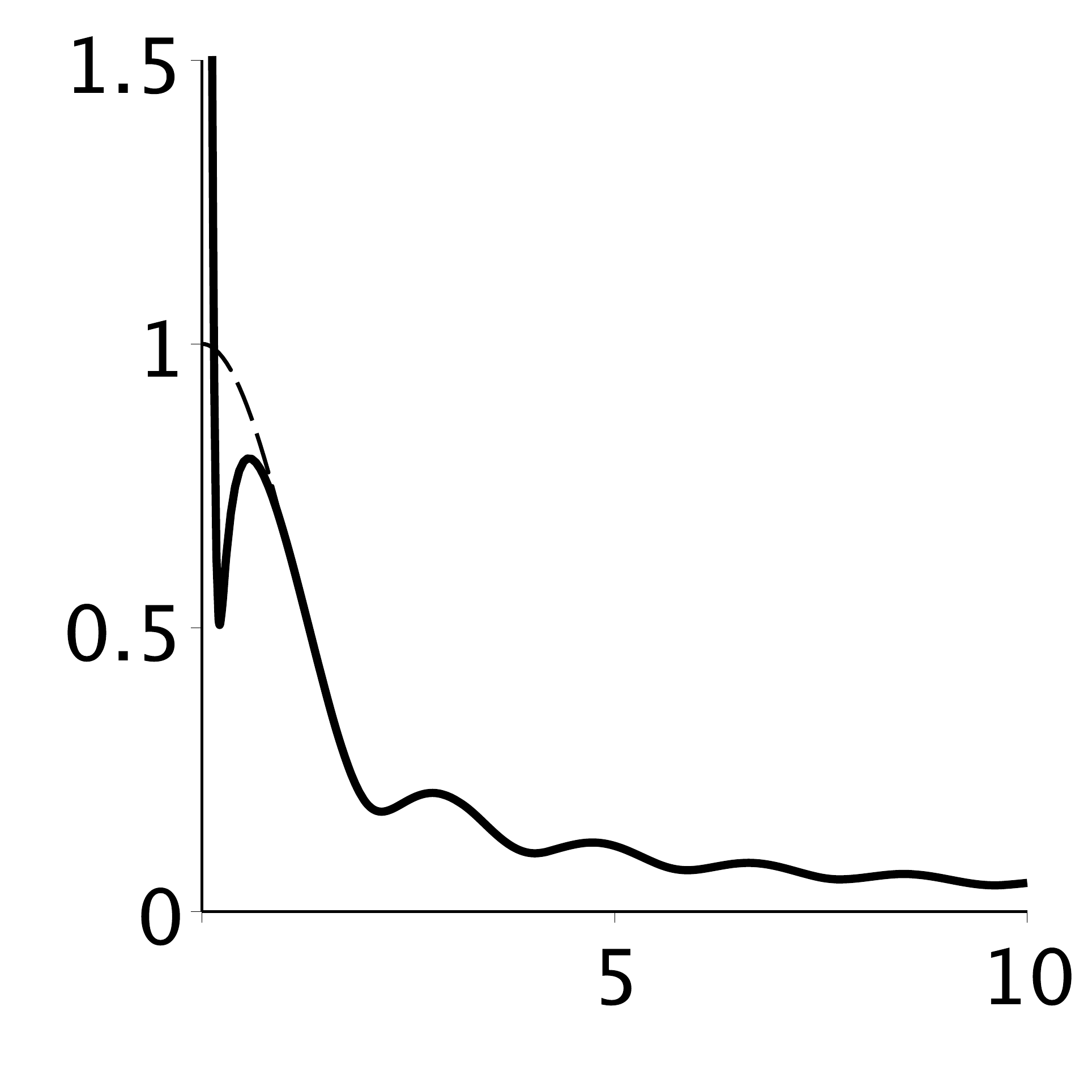}}%
    \put(2.2,1.87){\color[rgb]{0,0,0}\makebox(0,0)[lb]{ { $D=4,\alpha=4$} }}%
    \put(1.8,1.5){\color[rgb]{0,0,0}\makebox(0,0)[lb]{ { $|\mathcal{M}_1|$} }}
    \put(2.34,0.99){\color[rgb]{0,0,0}\makebox(0,0)[lb]{ { $\sigma_\alpha/\hbar$} }}%
    
        \put(-0.1,-0.01){\includegraphics[width=1\unitlength]{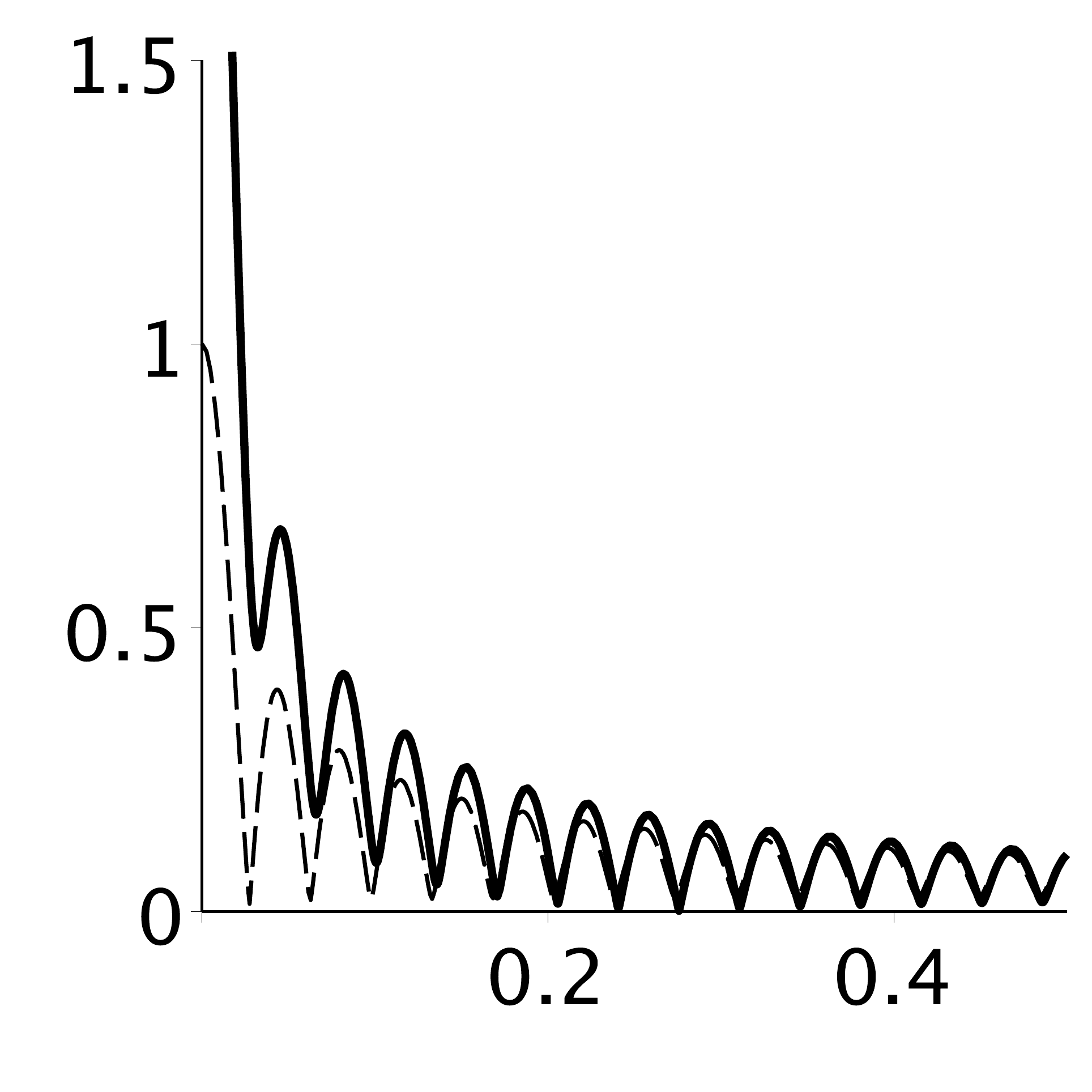}}%
    \put(0.2,0.87){\color[rgb]{0,0,0}\makebox(0,0)[lb]{ { $D=2,\alpha=10$} }}%
    \put(-0.2,0.5){\color[rgb]{0,0,0}\makebox(0,0)[lb]{ { $|\mathcal{M}_1|$} }}
    \put(0.34,-0.01){\color[rgb]{0,0,0}\makebox(0,0)[lb]{ { $\sigma_\alpha/\hbar$} }}%
        \put(0.9,-0.01){\includegraphics[width=1\unitlength]{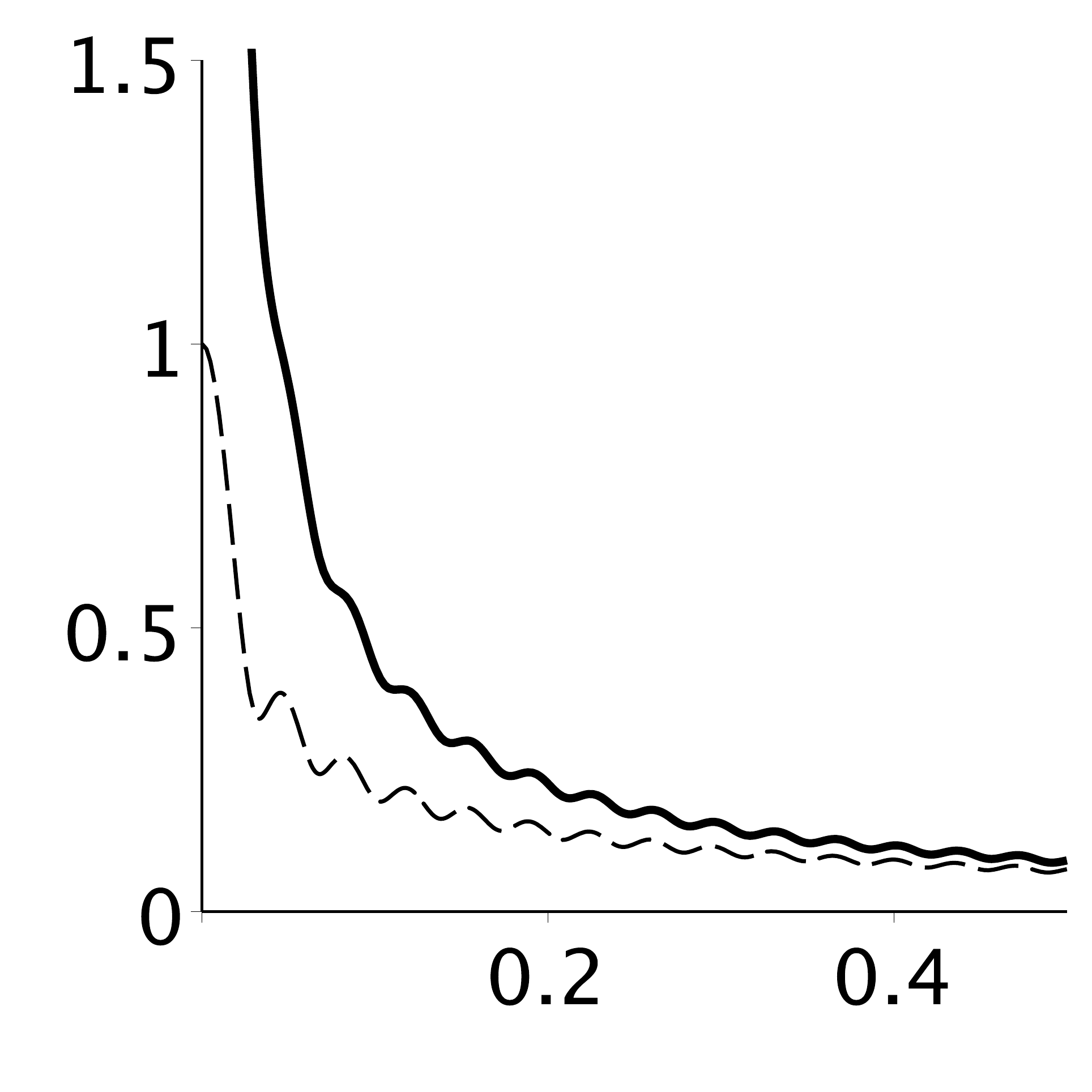}}%
    \put(1.2,0.87){\color[rgb]{0,0,0}\makebox(0,0)[lb]{ { $D=3,\alpha=10$} }}%'
    \put(0.8,0.5){\color[rgb]{0,0,0}\makebox(0,0)[lb]{ { $|\mathcal{M}_1|$} }}
    \put(1.34,-0.01){\color[rgb]{0,0,0}\makebox(0,0)[lb]{ { $\sigma_\alpha/\hbar$} }}%
        \put(1.9,-0.01){\includegraphics[width=1\unitlength]{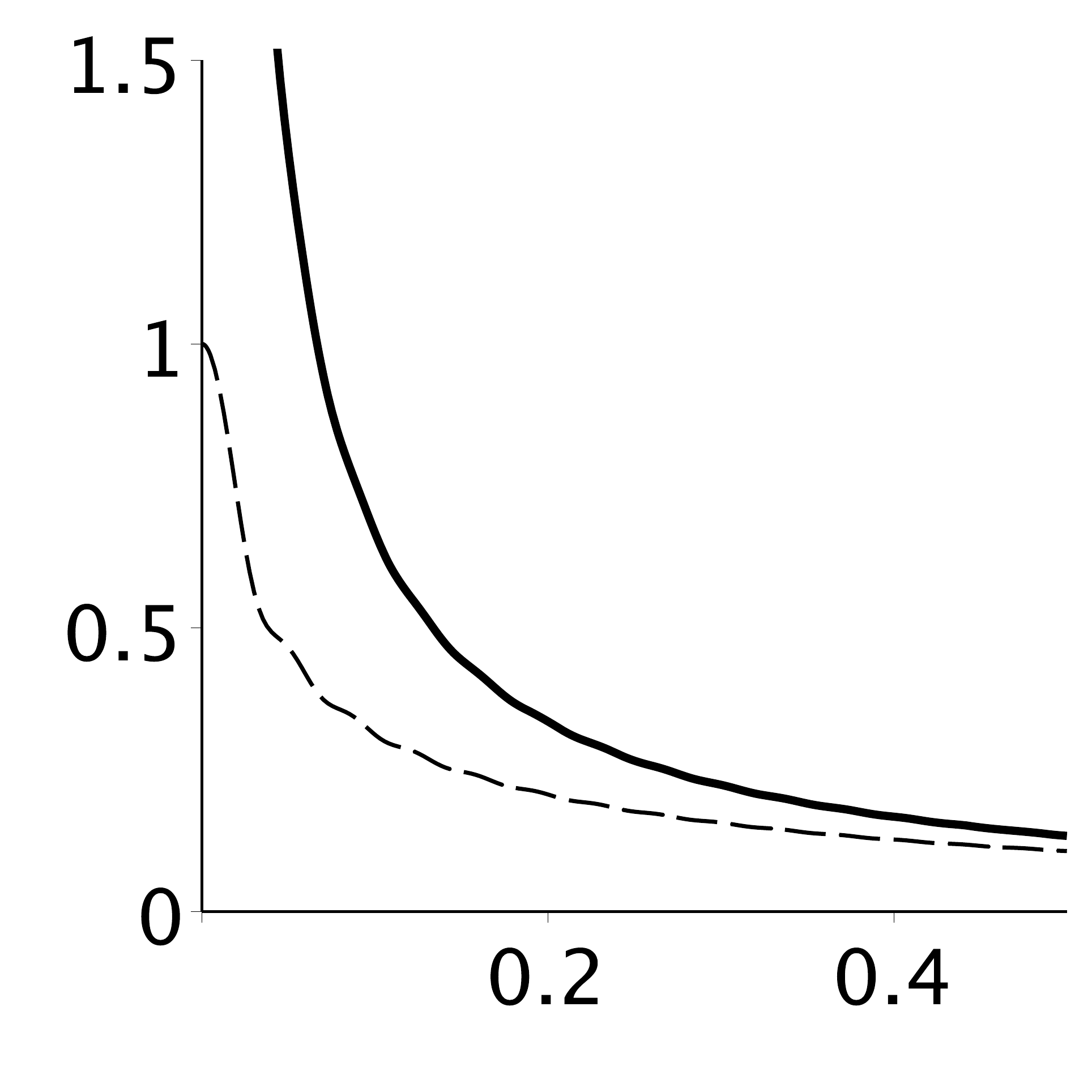}}%
    \put(2.2,0.87){\color[rgb]{0,0,0}\makebox(0,0)[lb]{ { $D=4,\alpha=10$} }}%
    \put(1.8,0.5){\color[rgb]{0,0,0}\makebox(0,0)[lb]{ { $|\mathcal{M}_1|$} }}
    \put(2.34,-0.01){\color[rgb]{0,0,0}\makebox(0,0)[lb]{ { $\sigma_\alpha/\hbar$} }}%
  \end{picture}%
\caption{Numerical evaluation of the modulation factor from the stationary phase approximation. The panels shows $| \mathcal{M}_{k=1} |$ as a function of $\sigma_\alpha / \hbar$ for nine different cases of the dimension and the order of the perturbative potential (physical parameters are set to unity). Solid curves shows results from SPA (\ref{SPA_final}), while thin black dashed curves shows the modulation factors calculated numerically from (\ref{OneDimensionalIntegral}). We have observed that the modulation factors calculated from SPA are generally indistinguishable from the numerical results in the $\hbar \rightarrow 0$ limit.
For the two rows with the lowest perturbative potentials illustrated here ($\alpha=2,4$) this happens already during the first oscillation.
An exception is the $D=3, \ \alpha=2$ panel, which is one of the cases where SPA is exact (see table \ref{table_for_M_k}). The fact that $| \mathcal{M}_{k=1} |=0$ in each oscillation also signals a super-shell structure for this case, as observed in \cite{OgrenPRA2007, JoPA2005}. 
When the order of the perturbation is increased substantially ($\alpha=10$), the results  from SPA are no longer close to the exact even after several oscillations (lowest row).
\label{fig:ModulationFactor}
}
\end{figure}

The perturbative POT in use in this article is valid to leading order in $\hbar^{-1}$.
Therefore we promote an analytic alternative to numerical integration, in the cases where the integral (\ref{OneDimensionalIntegral}) 
can not be given explicitly, the stationary phase approximation (SPA) to leading order in $\hbar^{-1}$. 
A few such examples are evaluated numerically in figure \ref{fig:ModulationFactor}. 

We can rewrite the integral  (\ref{OneDimensionalIntegral}) onto a standard form for Fourier integrals ($\hbar= 1/\lambda   \rightarrow 0 $ in the classical limit), according to 
\begin{gather}
\mathcal{M}_k =(D-1)  e^{-i k \sigma_\alpha a_0 / \hbar} \int_0^1 f(\ell) e^{i \lambda h(\ell) } d\ell, \nonumber \\ f(\ell)=\ell^{D-2}, \ h(\ell) =  - k \sigma_\alpha \sum_{j=1}^{\lfloor\alpha/2\rfloor}a_{j} \ell^{2j}. \label{SPA_Fourier_Integral}
\end{gather}
It is now our purpose to discuss the asymptotic expansion ($\lambda \to \infty$) according to SPA \cite{wong}.
The leading order contributions of the integral in (\ref{SPA_Fourier_Integral}), normally comes from the stationary points $\ell_0 \in (0,1)$, i.e., for which 
\begin{equation}
\Delta S'(\ell) = - \sigma_\alpha \sum_{j=1}^{\lfloor\alpha/2\rfloor} 2j a_{j} \ell^{2j-1} = 0. 
\label{Delta_S_stationary_points}
\end{equation}
For the polynomials found in table \ref{table_for_Delta_S}, there are no stationary points within the interval $0< \ell_0 \leq 1$, while  for $\ell=0$ we trivially have $\Delta S'(\ell) =0$. Given that \eqref{Delta_S_conjecture} holds, this is true for all $\alpha$. Since $x_0 \in (-1,1)$ for all zeroes $x_0$ of the Legendre polynomials $P_\alpha(x)$, it follows from the Gauss-Lucas theorem that the zeros $z_0$ of $P'_\alpha(z)$ also satisfy $z_0 \in (-1,1) $, hence $P'_\alpha(1/\ell_0) = 0$ implies $\ell_0 \notin [-1,1]$. 
The zeros are also simple, i.e. $P''_\alpha(z_0) \neq 0$. Using the recurrence relations for the Legendre polynomials, we find that (\ref{Delta_S_stationary_points}) simplifies to
\begin{equation}
\Delta S'(\ell) =  - \sigma_\alpha \alpha \ell^{\alpha - 1} P_{\alpha} \left( \frac{1}{\ell} \right) - \sigma_\alpha \ell^{\alpha } P'_{\alpha} \left( \frac{1}{\ell} \right) = \sigma_\alpha \ell^{\alpha - 1} P'_{\alpha - 1} \left( \frac{1}{\ell} \right)= 0. 
\label{Delta_S_stationary_points_Legendre}
\end{equation}
Hence, the only stationary point within the interval of the integral $\ell \in [0,1]$ is $\ell_0=0$ coming from the factor $\ell^{\alpha - 1}$ in the last part of (\ref{Delta_S_stationary_points_Legendre}). 
This means we can focus the asymptotic approximation of the integral in (\ref{SPA_Fourier_Integral}) onto the boundary points $\ell=1$ ($\ell=0$), which gives so called upper- (and lower-) end-point corrections \cite{wong} $I_\ell$, such that
\begin{equation}
\mathcal{M}_k \simeq (D-1)  e^{-i k \sigma_\alpha a_0 / \hbar} \left( I_1 + I_0 \right) \label{SPAansatz}.
\end{equation} 
Let us stress that this situation is atypical for most potentials that are  treated within POT, where stationary points corresponds to so called rational tori \cite{bertab}.
However, it was confirmed for the three-dimensional ($D=3$) quartic perturbed ($\alpha=2$) HO in \cite{JoPA2005}, that the leading order contributions came from those end-point corrections also in the exact trace formula. The end-point corrections could then be interpreted as corresponding to the classical diameter- ($\ell=0$) and circular- ($\ell=1$) periodic orbits.

For the upper integration limit $\ell=1$ (maximal angular momenta), we have $ f(\ell)=1$ and $ h'(\ell) \neq 0  $ in (\ref{SPA_Fourier_Integral}), such that the upper end-point contributes with a term \cite{wong}
\begin{equation}
I_1 \equiv -  \frac{i  f(1)}{\lambda h'(1)} e^{i\lambda  h(1)}   =\frac{ i    \hbar }{k  \sigma_\alpha \sum_{j=1}^{\lfloor\alpha/2\rfloor} 2j a_{j}} e^{-i k \sigma_\alpha \sum_{j=1}^{\lfloor\alpha/2\rfloor}a_{j}/\hbar}. \label{SPAI1}
\end{equation}

The lower integration limit $\ell=0$ (minimal angular momenta) needs special attention for $D \geq 3$, since, first we then have $ f(\ell)=0$, secondly it is a stationary point, i.e., $ h'(\ell) =0 $.
In this case this lower end-point contributes with a leading order term 
\begin{equation}
I_0 \equiv\int_0^\infty \ell^{D-2} e^{-i k \sigma_\alpha a_1 \ell^2 / \hbar} d\ell = \frac{\Gamma \left( \frac{D-1}{2} \right) }{2}  \left(\frac{ \hbar}{  k \sigma_\alpha a_1}   \right)^{\frac{D-1}{2}}     e^{- i \left( D-1 \right) \frac{ \pi}{4} }.  \label{SPAI0}
\end{equation}
Hence, from (\ref{SPAansatz}), (\ref{SPAI1}), and (\ref{SPAI0}) we can finally conclude that the asymptotic form of the modulation factor as obtained from SPA for an arbitrary even monomial perturbation to a harmonic oscillator in $D \geq 2$ dimensions is
\begin{align}
 &\mathcal{M}_k \simeq   \left( D-1 \right) i \hbar e^{-i k \sigma_\alpha a_0 / \hbar}  \times \nonumber \\
&\left(  \left[\frac{e^{-i k \sigma_\alpha \sum_{j=1}^{\lfloor\alpha/2\rfloor}a_{j}/\hbar} }{k  \sigma_\alpha \sum_{j=1}^{\lfloor\alpha/2\rfloor} 2j a_{j}} + \mathcal{O} \left( \hbar \right)  \right]  +   \frac{ \Gamma \left( \frac{D-1}{2} \right) }{2\hbar} \left(\frac{ \hbar}{  k \sigma_\alpha a_1}   \right)^{\frac{D-1}{2}}    e^{-i   \left( D+1 \right)  \frac{ \pi}{4}} \right). \label{SPA_final}
\end{align}
We observe that the SPA gives the exact integral in the $D=3$ and $\alpha=2,3$ cases, see table \ref{table_for_M_k}, since then the only two terms are both of order $\hbar$.
For $D=2$ the circular orbit  ($ \propto e^{i k  \Delta S(1) / \hbar}  $) is suppressed by a factor $\sqrt{\hbar}$, while for $D\geq 4$ the diameter orbit ($ \propto e^{i k  \Delta S(0) / \hbar}  $) is suppressed.
From the exact integrals in table \ref{table_for_M_k} it is seen that the next to leading order $\hbar^{-1}$ corrections of the circular orbit terms dominate the $\hbar^{-1}$ order of the leading diameter term already for $D>5$.
Further on, we can see from the cases in table \ref{table_for_M_k}, that (\ref{SPA_final}) then seems to exactly reproduce the leading order circular term in any dimension, while the diameter term  from (\ref{SPA_final}) seems exact only in odd dimensions. 

More important, the formula (\ref{SPA_final}) is certainly not restricted only to $\alpha=2,3$, and we report on a few numerically investigated cases in figure \ref{fig:ModulationFactor}.

\subsection{Applications with radially symmetric polynomial perturbations} \label{PolynomialPerturbations}

Following the same procedure as leading from (\ref{eq:FirstOrderAction}) to (\ref{eq:DeltaS_as_a_function_of_L}) but for $N$ perturbative terms,
it is straightforward to consider the more general polynomial perturbations
\begin{equation}
\Delta H=\sum_{j=1}^{N}\varepsilon_{j}\left|\mathbf{q}\right|^{2\alpha_{j}}.
\end{equation}
We briefly discuss one such realistic example here, and hope future readers can apply it to different perturbations within their own field of study. Motivated by the
mean-field description of weakly interacting fermions in a
harmonic trap \cite{OgrenPRA2007, HeiselbergPRL2002}, the perturbation in $D=3$ dimensions
is here proportional to a mean-field interaction parameter $\left| U_0 \right|\ll1$
times the following particle density in the Thomas-Fermi approximation 
\begin{equation}
\rho_{TF} = \rho_0 \left(1- \frac{r^{2}}{R_{TF}^{2}}\right)^{3/2} \simeq \rho_0 \left( 1-\frac{3}{2} \frac{r^{2}}{R_{TF}^{2}} +\frac{3}{8} \frac{r^{4}}{R_{TF}^{4}} + \frac{1}{16} \frac{r^{6}}{R_{TF}^{6}}+... \right).
\label{TF_approximation_for_Fermi_gas}
\end{equation}
We then consider the perturbative semiclassical action for a HO, with a
modified trap frequency $\omega_{\textnormal{{eff}}}=\sqrt{\omega^{2}+3 U_0 \rho_0/R_{TF}^{2}}$ due to the second term in (\ref{TF_approximation_for_Fermi_gas}), according to the contributions from the two last terms in (\ref{TF_approximation_for_Fermi_gas})
\begin{align}
\Delta S=-\frac{ U_0 \rho_0}{16R_{TF}^{6}}\int_{0}^{\frac{2\pi}{\omega}} \biggl( &6R_{TF}^{2}\left[a^{2}\cos^{2}(\omega t)+b^{2}\sin^{2}(\omega t)\right]^{2}  \nonumber \\
&+\left[a^{2}\cos^{2}(\omega t)+b^{2}\sin^{2}(\omega t)\right]^{3} \biggr) dt.
\end{align}
Hence, from the linearity of the integral, we have using (\ref{eq:DeltaS_as_a_function_of_L}) that 
\begin{equation} 
\Delta S \left( \ell \right) =-\frac{U_0 \rho_0 \pi R_{0}^{4}}{32\omega R_{TF}^{4}}\left[36+\frac{5R_{0}^{2}}{R_{TF}^{2}}-\left(12+\frac{3R_{0}^{2}}{R_{TF}^{2}}\right)  \ell^2 \right]. \label{Delta_S_for_polynomial_perturbation}
\end{equation}
The fact that there is only one non-constant term in (\ref{Delta_S_for_polynomial_perturbation}),
is in agreement with an alternative perturbative semiclassical analysis for this mean-field potential performed using WKB wavefunctions \cite{Ogren, HeiselbergPRL2002}.
In particular this also means that the exact modulation factor is straightforward to obtain analytically in analogy to the case for $D=3$ in table \ref{table_for_M_k}.
Finally we stress that similar polynomial perturbations can be constructed (e.g.) with the help of table \ref{table_for_Delta_S}.

\section{Final trace formulae}

Combining (\ref{eq:PerturbativeTraceFormula}) and (\ref{OneDimensionalIntegral}) we can generally write the exact perturbative trace formula on the following compact  form
\begin{equation}
g_{\mathrm{pert}}(E) \simeq \frac{E^{D-1}}{\left(D-2\right)! \left(\hbar\omega\right)^{D} }\,\text{Re}\left\{ \sum_{k=-\infty}^{\infty}(-1)^{Dk}  \int_0^1 \ell^{D-2} e^{ - i k \sigma_\alpha \ell^{\alpha} P_\alpha ( \frac{1}{\ell} ) /\hbar} d\ell \ e^{ i k S_0/\hbar}\right\},  \label{GeneralFinalTF}
\end{equation}
where $S_0 = 2\pi E/\omega$, and the role of the order $\alpha$ of the perturbative potential $\varepsilon r^{2 \alpha}$ enters through the polynomial in the exponent, see table \ref{table_for_Delta_S} for examples.
For quartic- and sextic-perturbations ($\alpha=2,3$), the one-dimensional Fourier integral in (\ref{GeneralFinalTF}) can be expressed by the generalised hypergeometric function of (\ref{eq:HyperModulationfactor}), see table~\ref{table_for_M_k} for examples.

Finally, in all cases we can approximate the Fourier integral in (\ref{GeneralFinalTF}) with SPA (\ref{SPA_final}) such that the modulation factor only contains the leading order $\hbar^{-1}$ contributions for the diameter- and circular-orbits respectively in elementary functions.

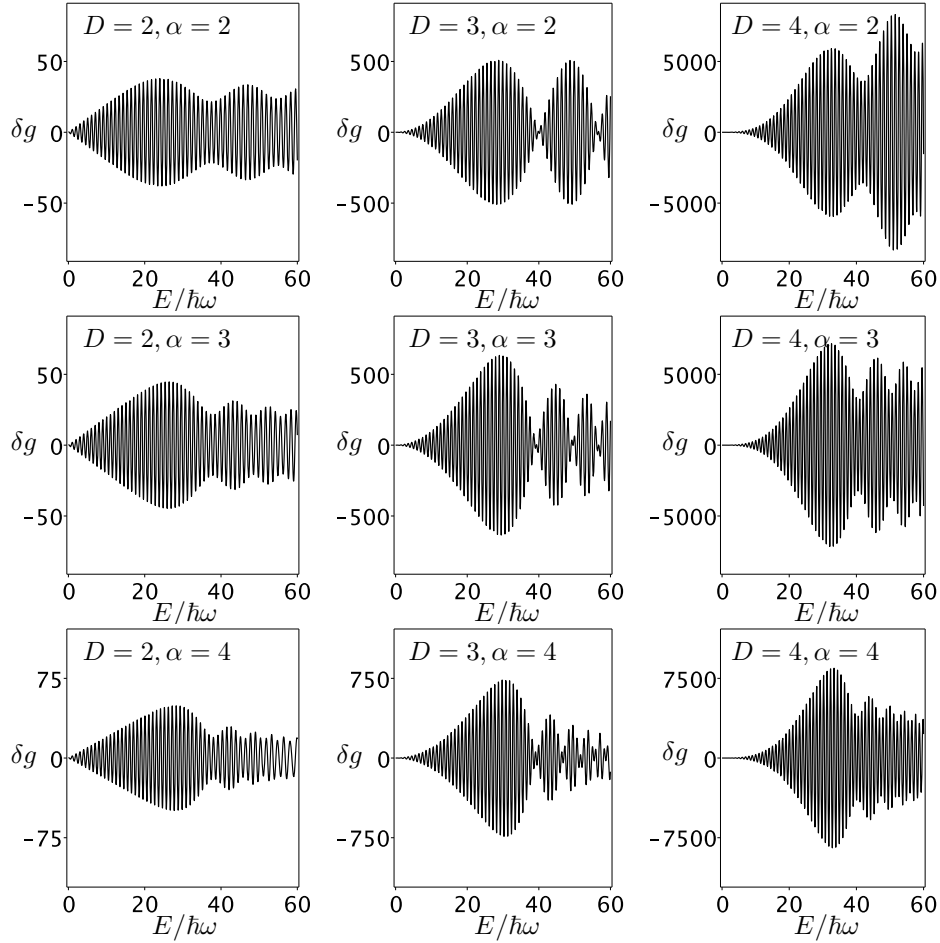
\begin{figure}
\centering{}
    \setlength{\unitlength}{118bp}%
  \begin{picture}(3,3)%
    	\put(-0.1,2){\includegraphics[width=1\unitlength]{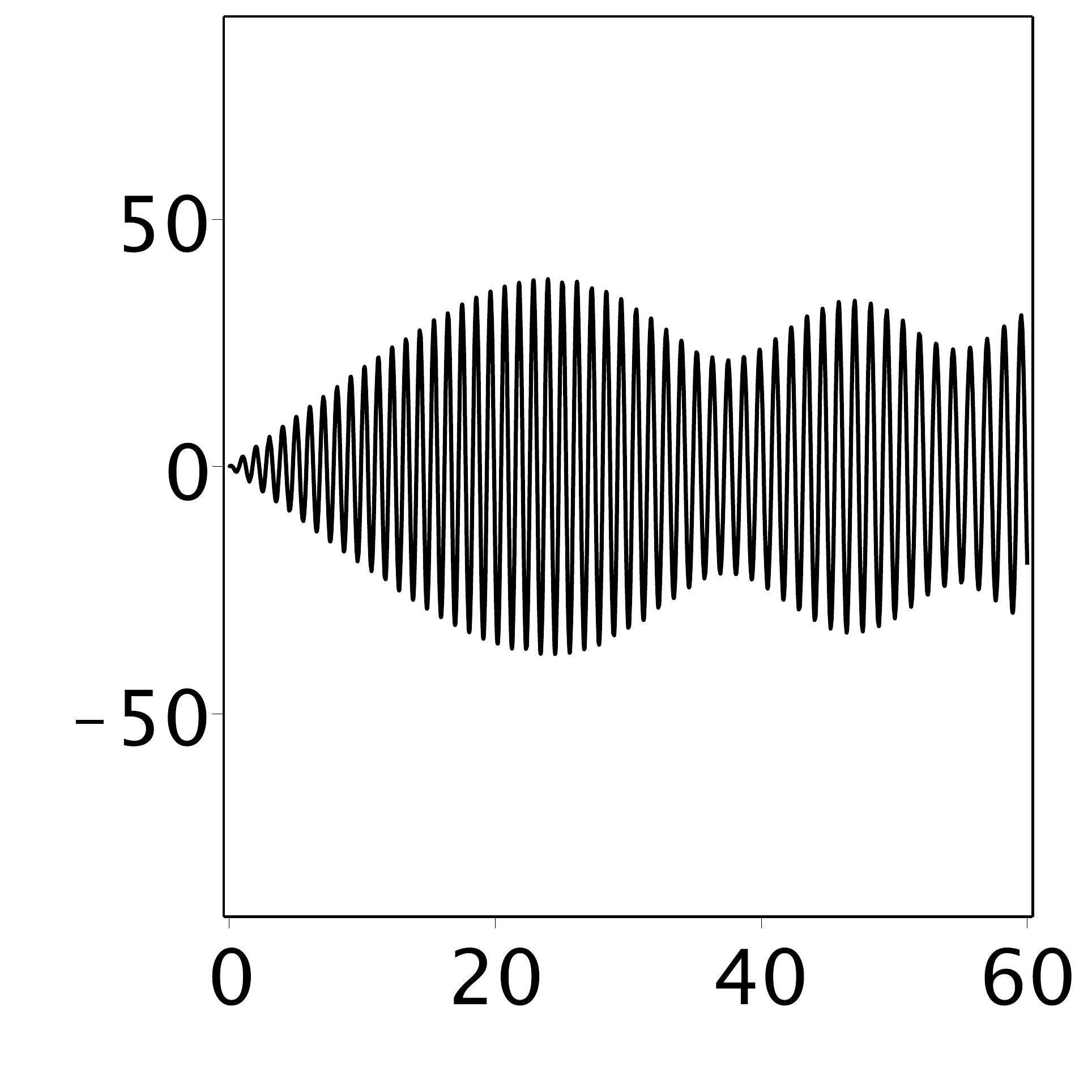}}%
    \put(0.1,2.87){\color[rgb]{0,0,0}\makebox(0,0)[lb]{ { $D=2,\alpha=2$} }}%
    \put(-0.13,2.54){\color[rgb]{0,0,0}\makebox(0,0)[lb]{ { $\delta g$} }}
    \put(0.32,1.99){\color[rgb]{0,0,0}\makebox(0,0)[lb]{ { $E/\hbar \omega$} }}%
    	\put(0.9,2){\includegraphics[width=1\unitlength]{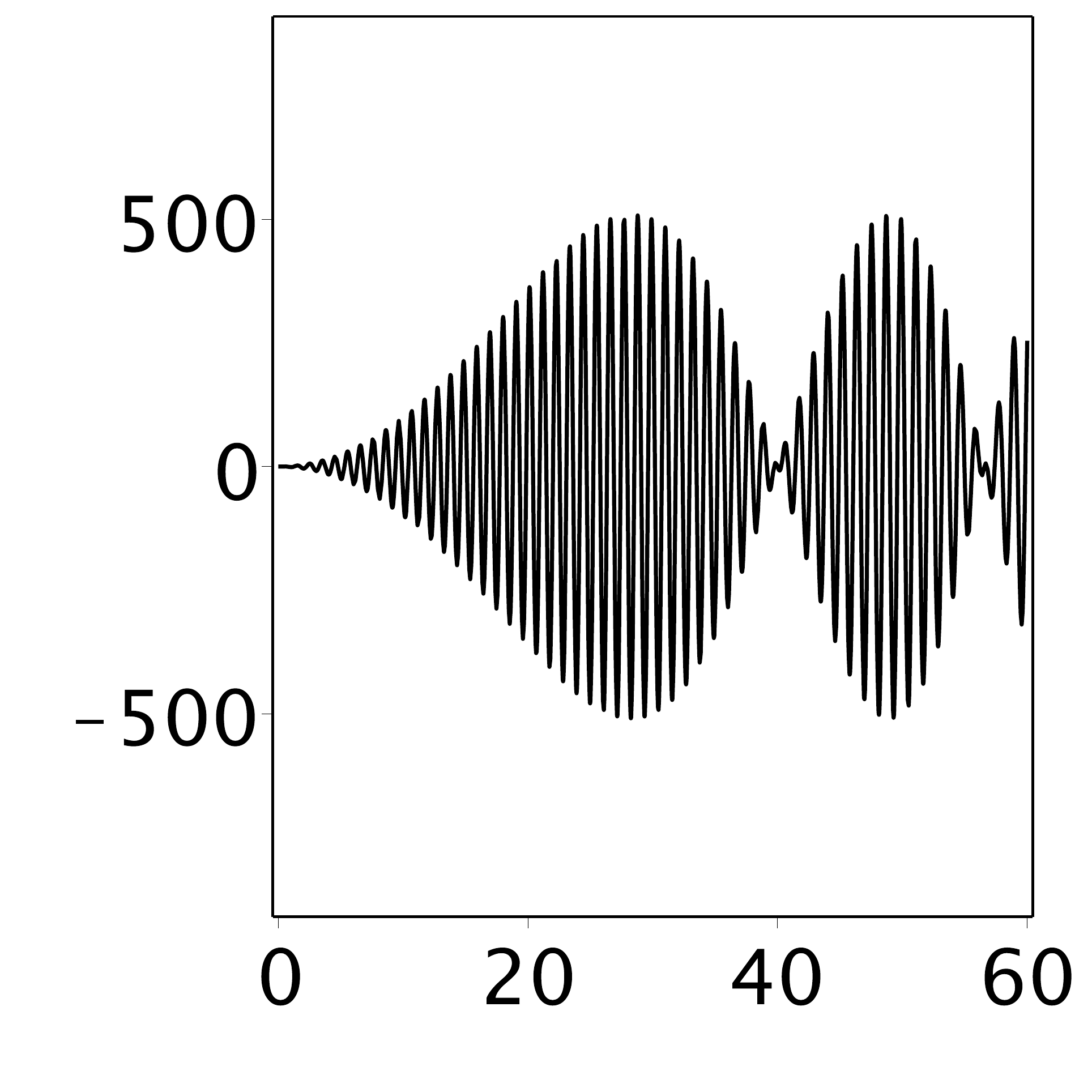}}%
    \put(1.14,2.87){\color[rgb]{0,0,0}\makebox(0,0)[lb]{ { $D=3,\alpha=2$} }}%        
    \put(0.91,2.54){\color[rgb]{0,0,0}\makebox(0,0)[lb]{ { $\delta g$} }}
    \put(1.35,1.99){\color[rgb]{0,0,0}\makebox(0,0)[lb]{ { $E/\hbar \omega$} }}%
    	\put(1.9,2){\includegraphics[width=1\unitlength]{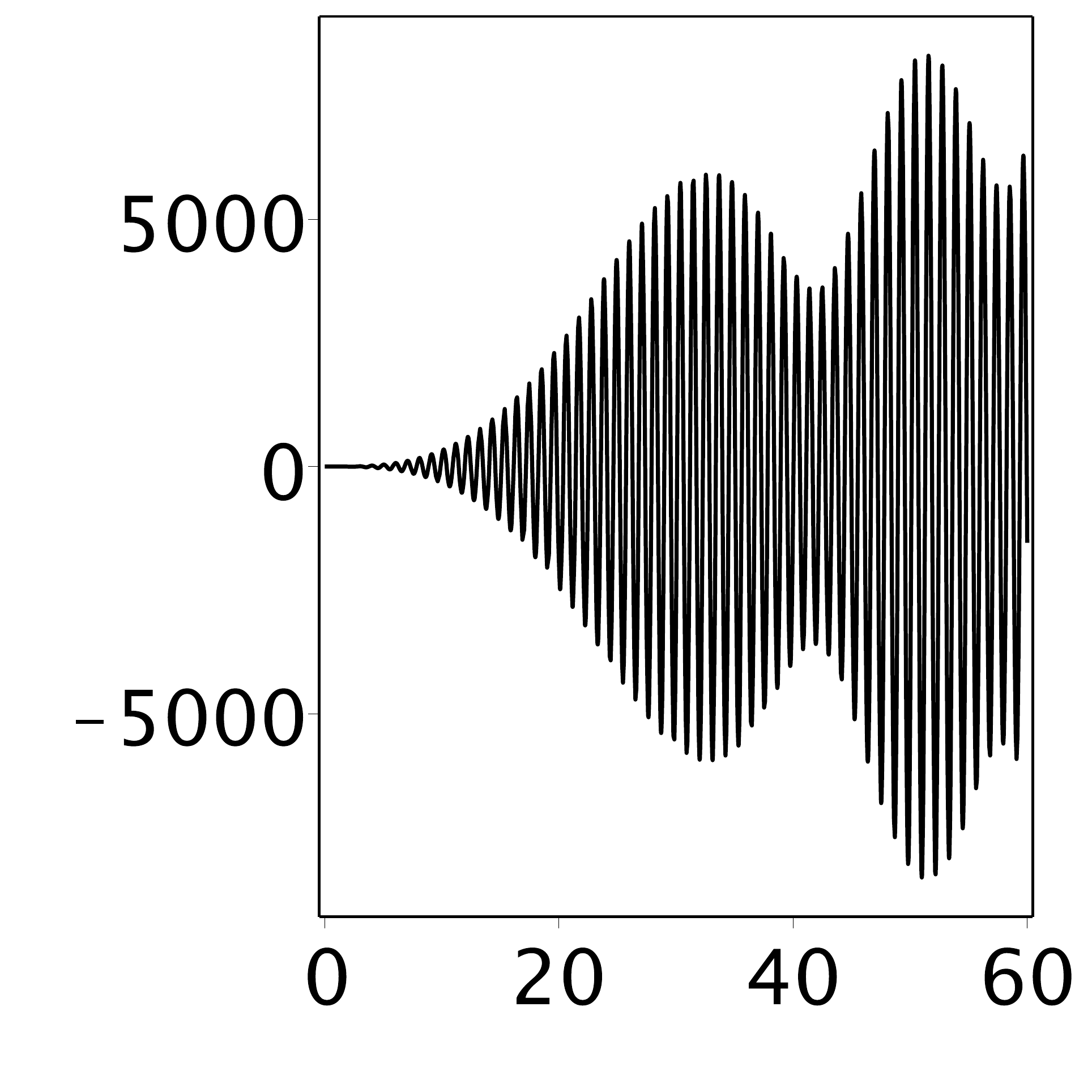}}%
    \put(2.17,2.87){\color[rgb]{0,0,0}\makebox(0,0)[lb]{ { $D=4,\alpha=2$} }}%
    \put(1.95,2.54){\color[rgb]{0,0,0}\makebox(0,0)[lb]{ { $\delta g$} }}
    \put(2.37,1.99){\color[rgb]{0,0,0}\makebox(0,0)[lb]{ { $E/\hbar \omega$} }}%    

    	\put(-0.1,1){\includegraphics[width=1\unitlength]{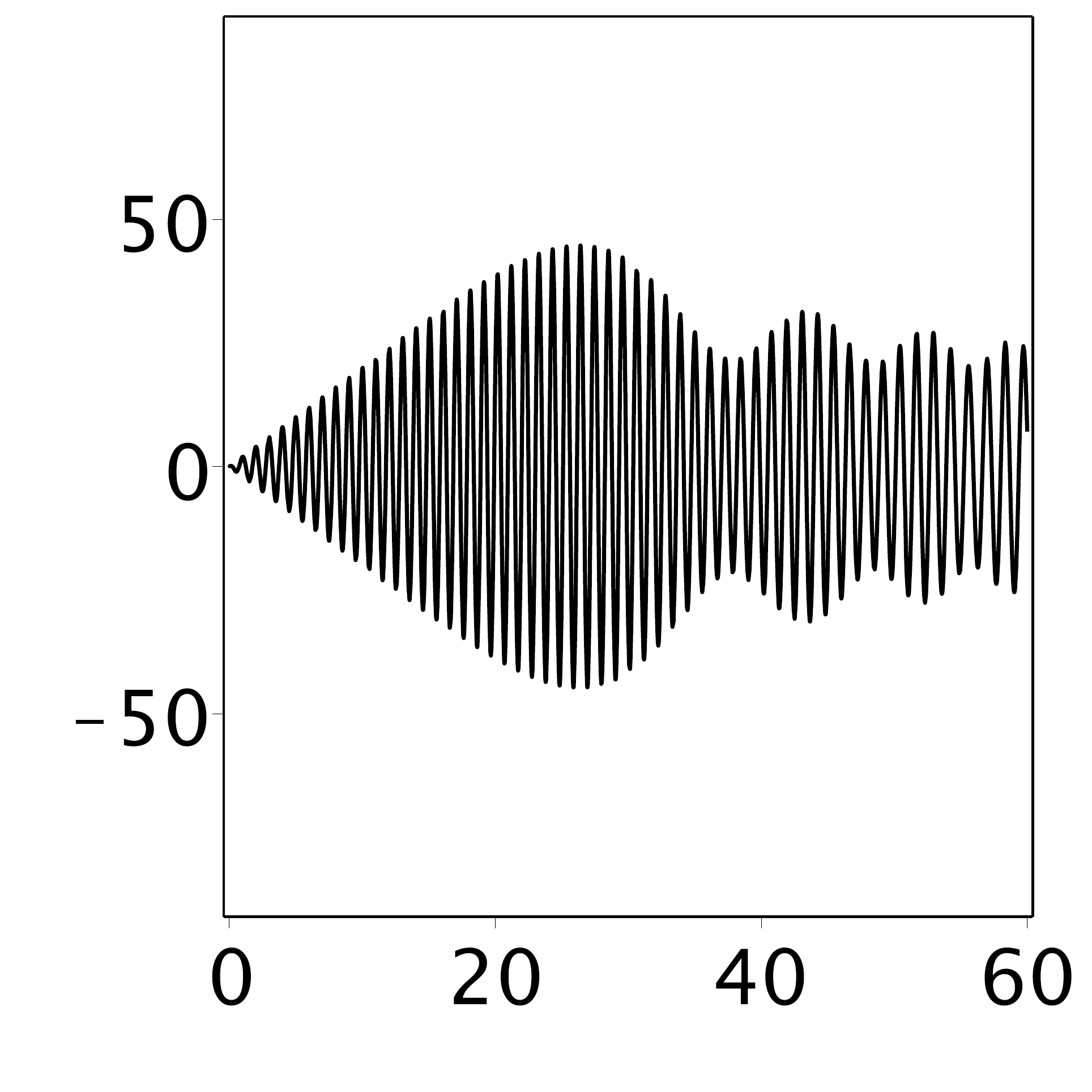}}%    	
    \put(0.1,1.87){\color[rgb]{0,0,0}\makebox(0,0)[lb]{ { $D=2,\alpha=3$} }}%    
    \put(-0.13,1.54){\color[rgb]{0,0,0}\makebox(0,0)[lb]{ { $\delta g$} }}
    \put(0.32,0.99){\color[rgb]{0,0,0}\makebox(0,0)[lb]{ { $E/\hbar \omega$} }}%
        \put(0.9,1){\includegraphics[width=1\unitlength]{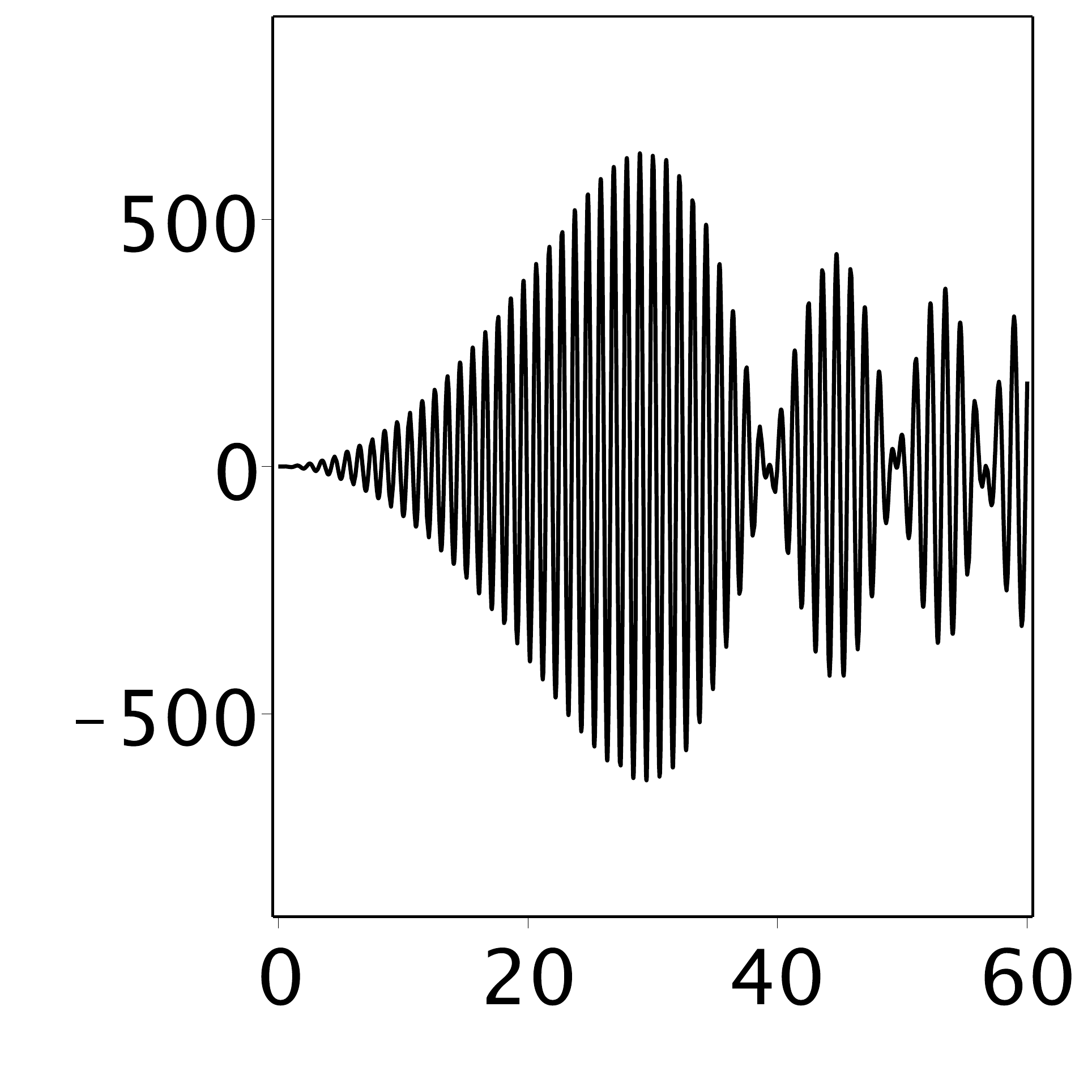}}%
    \put(1.14,1.87){\color[rgb]{0,0,0}\makebox(0,0)[lb]{ { $D=3,\alpha=3$} }}%
    \put(0.91,1.54){\color[rgb]{0,0,0}\makebox(0,0)[lb]{ { $\delta g$} }}
    \put(1.35,0.99){\color[rgb]{0,0,0}\makebox(0,0)[lb]{ { $E/\hbar \omega$} }}%
        \put(1.9,1){\includegraphics[width=1\unitlength]{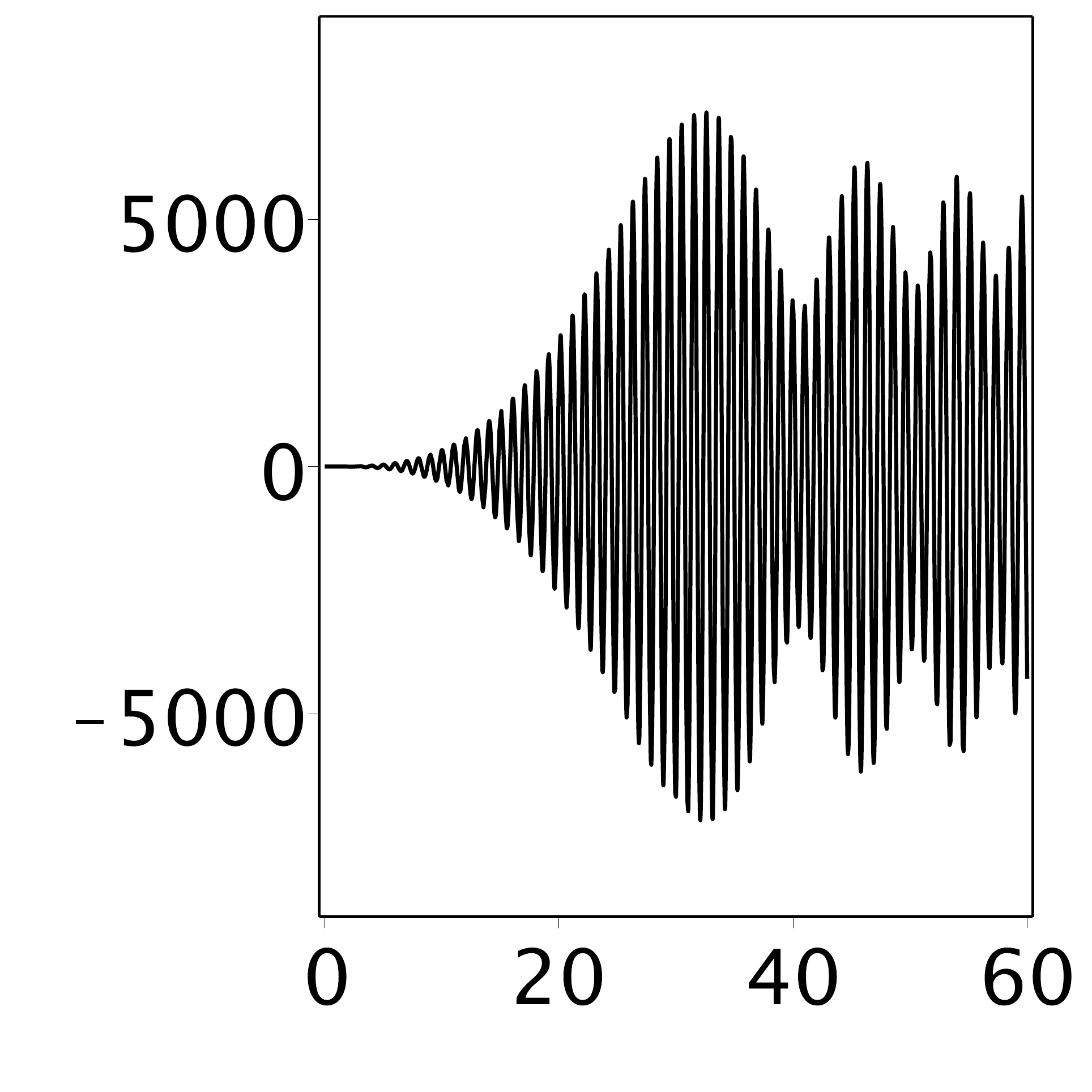}}%
    \put(2.17,1.87){\color[rgb]{0,0,0}\makebox(0,0)[lb]{ { $D=4,\alpha=3$} }}%
    \put(1.95,1.54){\color[rgb]{0,0,0}\makebox(0,0)[lb]{ { $\delta g$} }}
    \put(2.37,0.99){\color[rgb]{0,0,0}\makebox(0,0)[lb]{ { $E/\hbar \omega$} }}%    
    
        \put(-0.1,0){\includegraphics[width=1\unitlength]{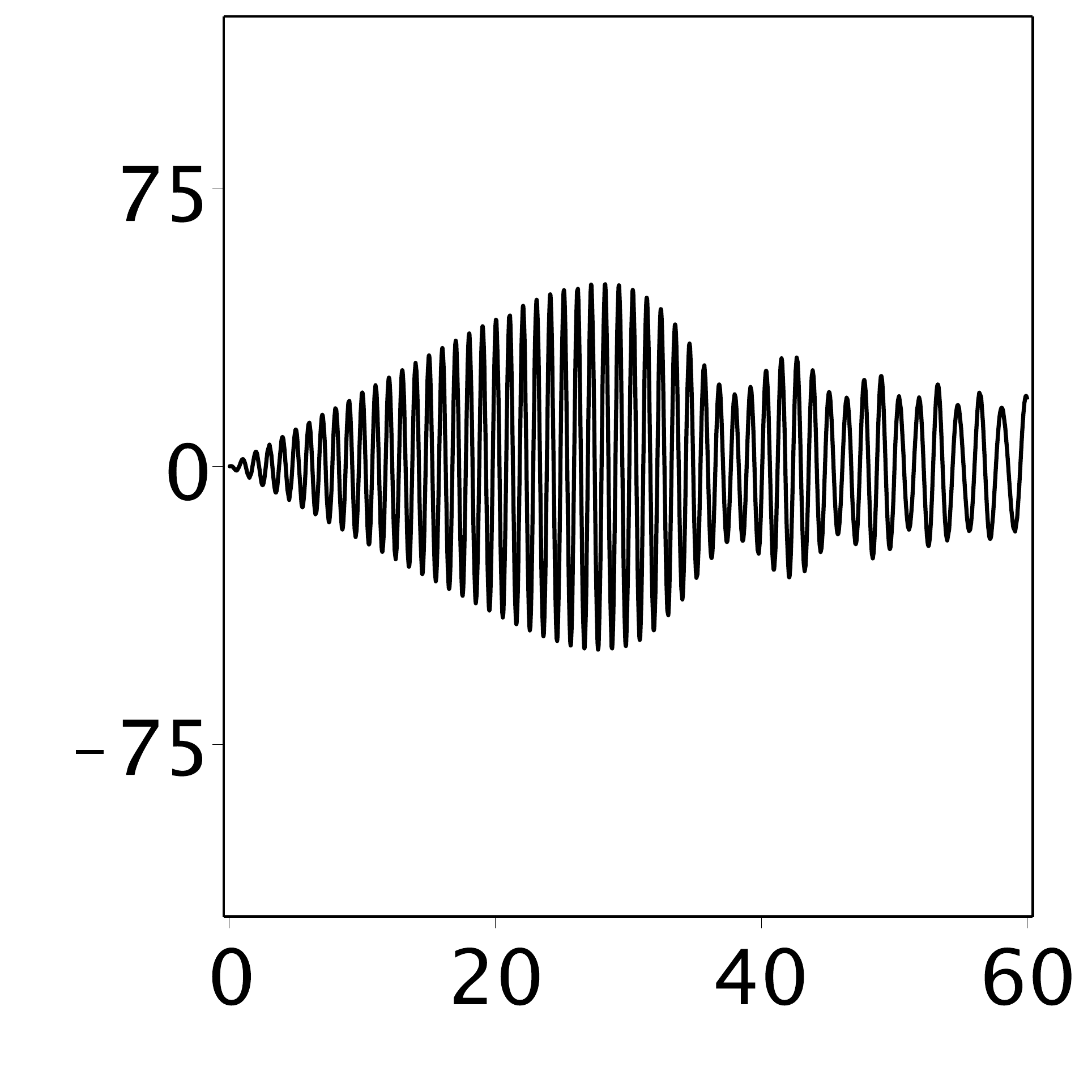}}%
    \put(0.1,0.87){\color[rgb]{0,0,0}\makebox(0,0)[lb]{ { $D=2,\alpha=4$} }}%
    \put(-0.13,0.54){\color[rgb]{0,0,0}\makebox(0,0)[lb]{ { $\delta g$} }}
    \put(0.32,-0.01){\color[rgb]{0,0,0}\makebox(0,0)[lb]{ { $E/\hbar \omega$} }}%
        \put(0.9,0){\includegraphics[width=1\unitlength]{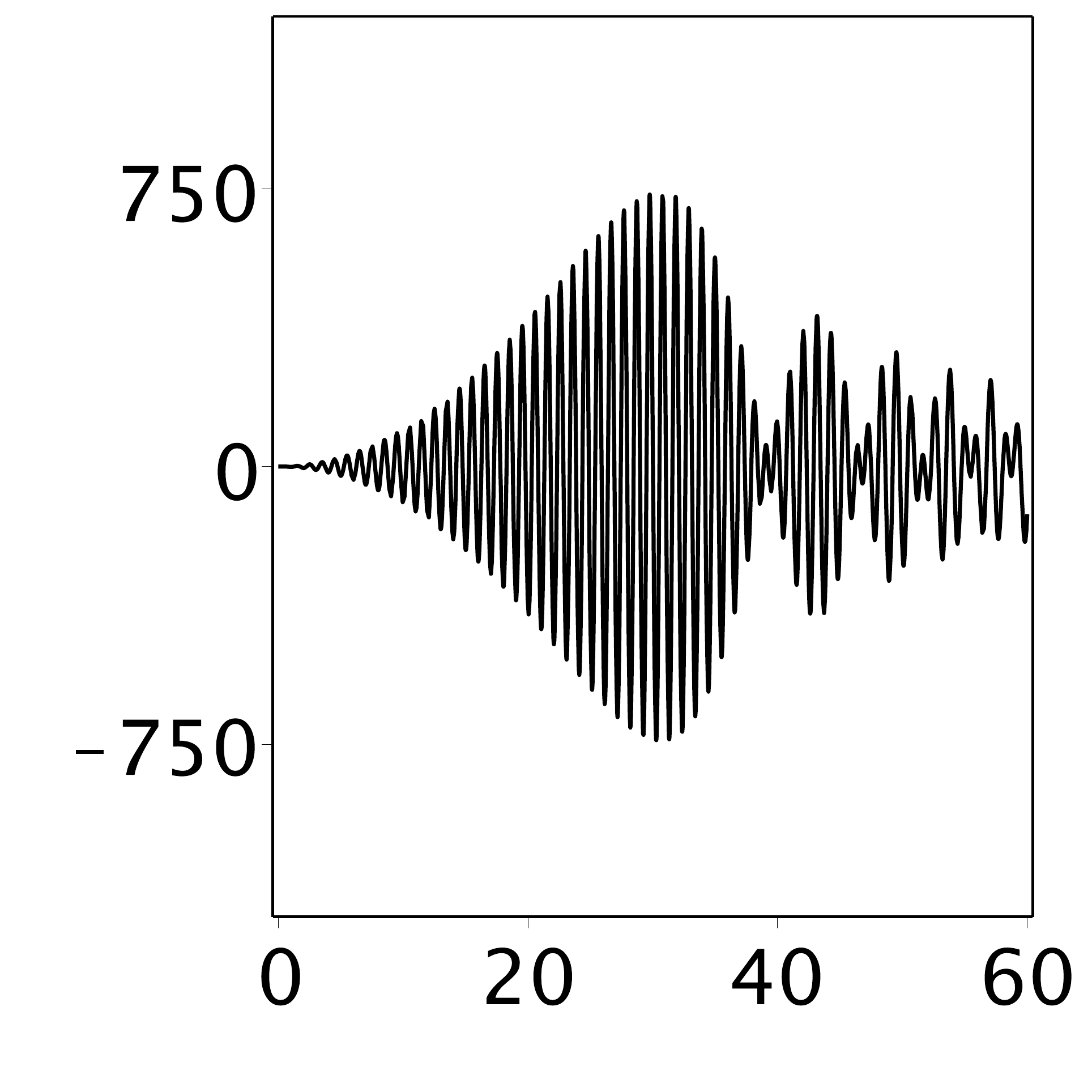}}%
    \put(1.14,0.87){\color[rgb]{0,0,0}\makebox(0,0)[lb]{ { $D=3,\alpha=4$} }}%
    \put(0.91,0.54){\color[rgb]{0,0,0}\makebox(0,0)[lb]{ { $\delta g$} }}
    \put(1.35,-0.01){\color[rgb]{0,0,0}\makebox(0,0)[lb]{ { $E/\hbar \omega$} }}%
        \put(1.9,0){\includegraphics[width=1\unitlength]{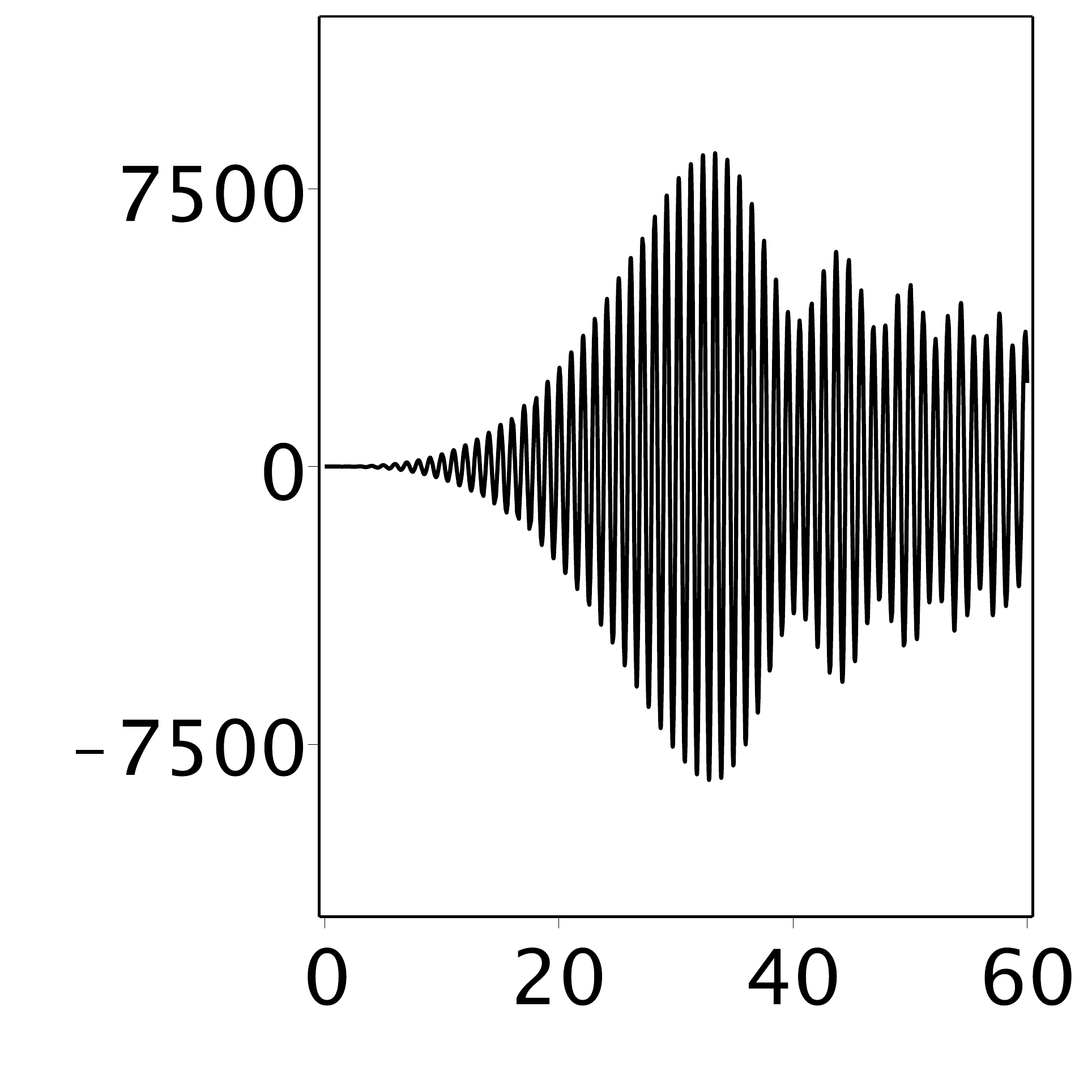}}%
    \put(2.17,0.87){\color[rgb]{0,0,0}\makebox(0,0)[lb]{ { $D=4,\alpha=4$} }}%
    \put(1.95,0.55){\color[rgb]{0,0,0}\makebox(0,0)[lb]{ { $\delta g$} }}
    \put(2.37,-0.01){\color[rgb]{0,0,0}\makebox(0,0)[lb]{ { $E/\hbar \omega$} }}%
  \end{picture}%

\caption{Numerical illustration of the trace formulae for the gross structure of the oscillating part of the density of states. 
The panels shows the $|k|=1$ terms of $\delta g_{\mathrm{pert}}$ as a function of $E/\hbar \omega$, calculated from (\ref{GeneralFinalTF}) with a Gaussian quadrature which is effective for moderate values of $\sigma_\alpha / \hbar$.
For the two cases $D=3$ and $\alpha=2,3$, we have in addition confirmed the validity of (\ref{SSS_TF}). 
In those two panels the analytic super-shell nodes given by (\ref{n_S_2}) and (\ref{n_S_3}) predicts the strength of the perturbation to be used in the $\alpha=2,3$ rows, for (e.g.) $n_s = 40$, to be $\varepsilon = 1.25\cdot 10^{-3}$ and $\varepsilon = 1.1 \cdot 10^{-5}$, respectively. For the last ($\alpha = 4$) row we chose $\varepsilon = 1.25 \cdot 10^{-7}$.
A local numerical investigation showed that only the two cases $D=3$ and $\alpha=2,3$ have prominent super-shell structure, i.e., where the amplitude of the envelope disappears in the super-shell nodes, while in for example the panel $D=3$ and $\alpha=4$ a tiny amplitude of the envelope remains (compare with the central panel of figure \ref{fig:ModulationFactor}).
We observe that the amplitudes of the shell oscillations are increasing by approximately a factor of ten when the spatial dimension is increased.
\label{fig:SSS}
}
\end{figure}

\subsection{Super-shell structures}

The cases $D=3$ and $\alpha=2,3$ (including the polynomial application discussed in section \ref{PolynomialPerturbations}) is special, since, according to table~\ref{table_for_M_k}, the modulation factor then only have two terms, both of the same order in $\hbar$. 
We now use the $D=3$ case of table \ref{table_for_M_k}  to calculate $\text{Re}\left\{ \sum_{k=-\infty}^{\infty}(-1)^{k}\mathcal{M}_{k}e^{ i k S_0/\hbar}\right\}$ from \eqref{eq:PerturbativeTraceFormula}, for which we  obtain (excluding the $k=0$ term)
\begin{equation}
\frac{2\hbar}{\sigma_\alpha a_1} \sum_{k=1}^{\infty} \frac{(-1)^{k}}{k} \left\{   \sin \left( \frac{k}{\hbar}\left[  S_0 - \sigma_\alpha  a_0 \right]  \right)   - \sin \left( \frac{k}{\hbar}\left[  S_0 - \sigma_\alpha \left( a_0+a_1  \right) \right]  \right)   \right\}. 
\end{equation}
As first reported in \cite{OgrenPRA2007} (for $\alpha=2$ and with a spin-factor of $2$), this allows us to use trigonometric identities to write the trace formula for the oscillating part of the DOS on a factorised form
\begin{equation}
\delta g_{\mathrm{pert}} \left( E \right) \simeq  \frac{ \omega^{2(\alpha-1)} E^{2-\alpha}}{ \pi \varepsilon a_1 \hbar^2}   \sum_{k=1}^{\infty} \frac{(-1)^{k}}{k}    \cos \left( \frac{k}{\hbar} \left[  S_0 - \sigma_\alpha  \left( a_0+\frac{a_1}{2}  \right) \right]  \right)   \sin \left( \frac{k\sigma_\alpha a_1}{2\hbar} \right) . \label{SSS_TF}
\end{equation}
From (\ref{Delta_S_for_alpha_2}) and (\ref{Delta_S_for_alpha_3}) we see that $2a_0=3$($5$) and $2a_1=-1$($-3$) for $\alpha=2$($3$). In both cases the dimension of (\ref{SSS_TF}) is $E^{-1}$ as it should for the DOS.
It is clear from the second factor in (\ref{SSS_TF}) that we have a prominent super-shell structure here, with so called super-shell nodes (i.e., where the envelope of $\delta g \left( E \right)$ is zero) when the argument of the sine is a multiple $s=1,2, \hdots$ of $\pi$, i.e., with the super-shell nodes $n_S$ (main HO quantum number) given for $\alpha=2$ by \cite{Ogren}
\begin{equation}
n_s  = \frac{E}{ \hbar \omega} = \sqrt{ \frac{2 s  \omega^3 }{ |\varepsilon| \hbar} },
\label{n_S_2}
\end{equation}
and for $\alpha=3$ by 
\begin{equation}
n_s  =\frac{E}{ \hbar \omega} = \sqrt[3]{\frac{2 s  \omega^4 }{3 |\varepsilon| \hbar^2} }.
\label{n_S_3}
\end{equation}

In figure \ref{fig:SSS} we illustrate the super-shell structure, and in particular the super-shell nodes (\ref{n_S_2}) and (\ref{n_S_3}), for the ($D=3$) cases $\alpha=2,3$, as opposed to (e.g.) the case $\alpha=4$. Be aware that using the SPA for small values of $\sigma_\alpha / \hbar$ can also generate false super-shell nodes, e.g., for $D=4$ and $\alpha=2$ (compare upper-right panel of figure \ref{fig:ModulationFactor}).
Let us finally stress that the results presented in the two panels $D=2,3$ and $\alpha=2$ in figure \ref{fig:SSS} agrees with earlier work published in \cite{crpert} and \cite{JoPA2005} respectively.
In the latter $D=3$ case the validity of the analytic result presented here have then implicitly also been checked against the DOS calculated numerically from the corresponding Schr\"{o}dinger equation \cite{JoPA2005}.
It is important to mention that due to the restriction in the orbits included in the perturbative trace formula, it does not converge to the full semiclassical (EBK) spectrum for spherical systems, i.e., where individual energy levels can be labelled by two quantum numbers. It rather gives the smooth DOS within each energyband of the main HO quantum shells (see figure~\ref{fig:EBK} of the next section), and it marks the start- ($l_{\mathrm{min}}$) and end-point ($l_{\mathrm{max}}$) of such a band \cite{Ogren}. 

\subsection{Comparison with the density of states from EBK theory}

\begin{figure}
\centering{}
    \setlength{\unitlength}{118bp}%
  \begin{picture}(3,2.4)%
    	\put(-0.05,1.8){\includegraphics[width=1\unitlength]{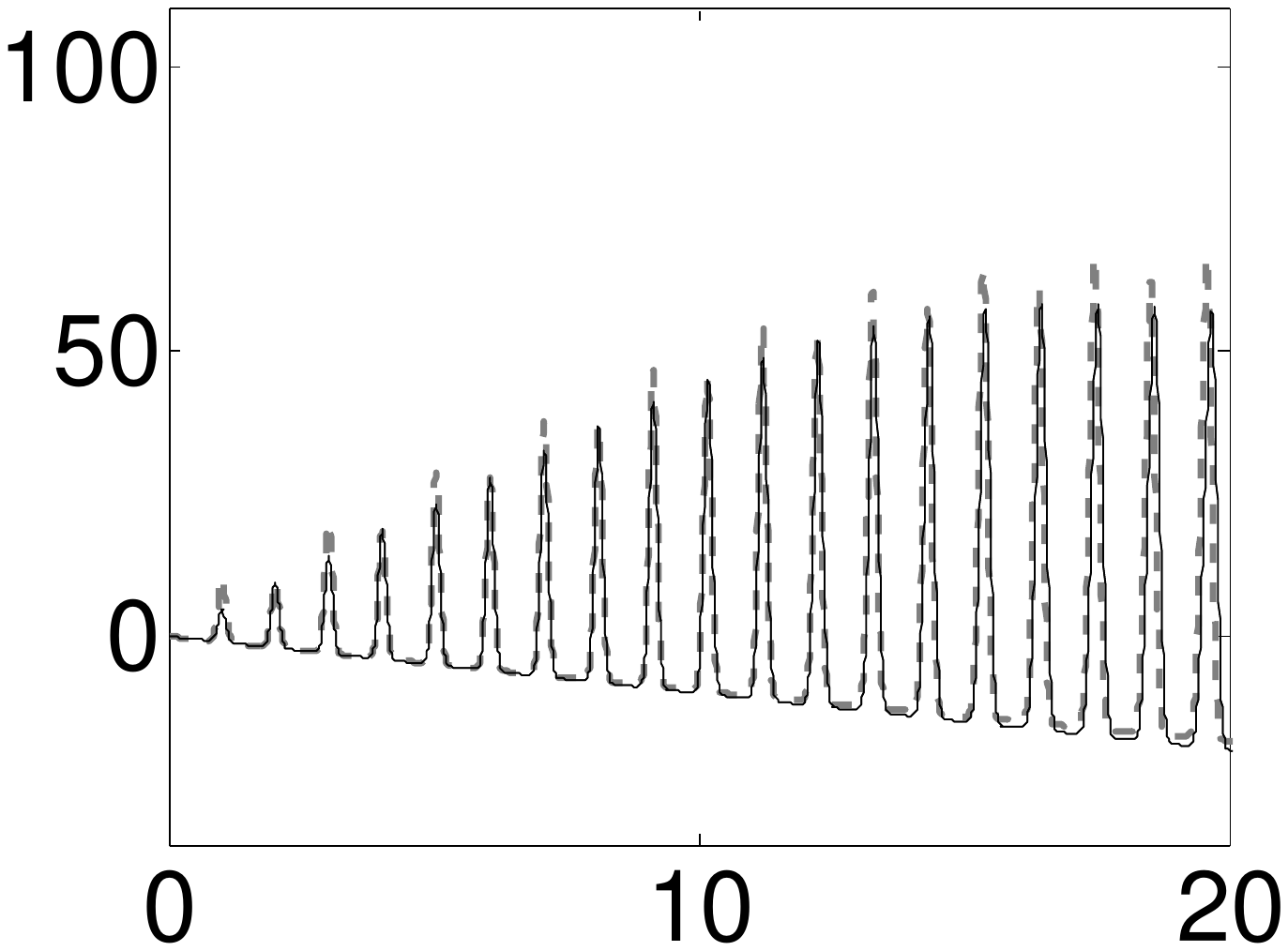}}%
    \put(0.1,2.37){\color[rgb]{0,0,0}\makebox(0,0)[lb]{ { $D=2,\alpha=2$} }}%
    \put(-0.13,2.1){\color[rgb]{0,0,0}\makebox(0,0)[lb]{ { $\delta g$} }}
    \put(0.32,1.7){\color[rgb]{0,0,0}\makebox(0,0)[lb]{ { $E/\hbar \omega$} }}%
    	\put(0.95,1.8){\includegraphics[width=1\unitlength]{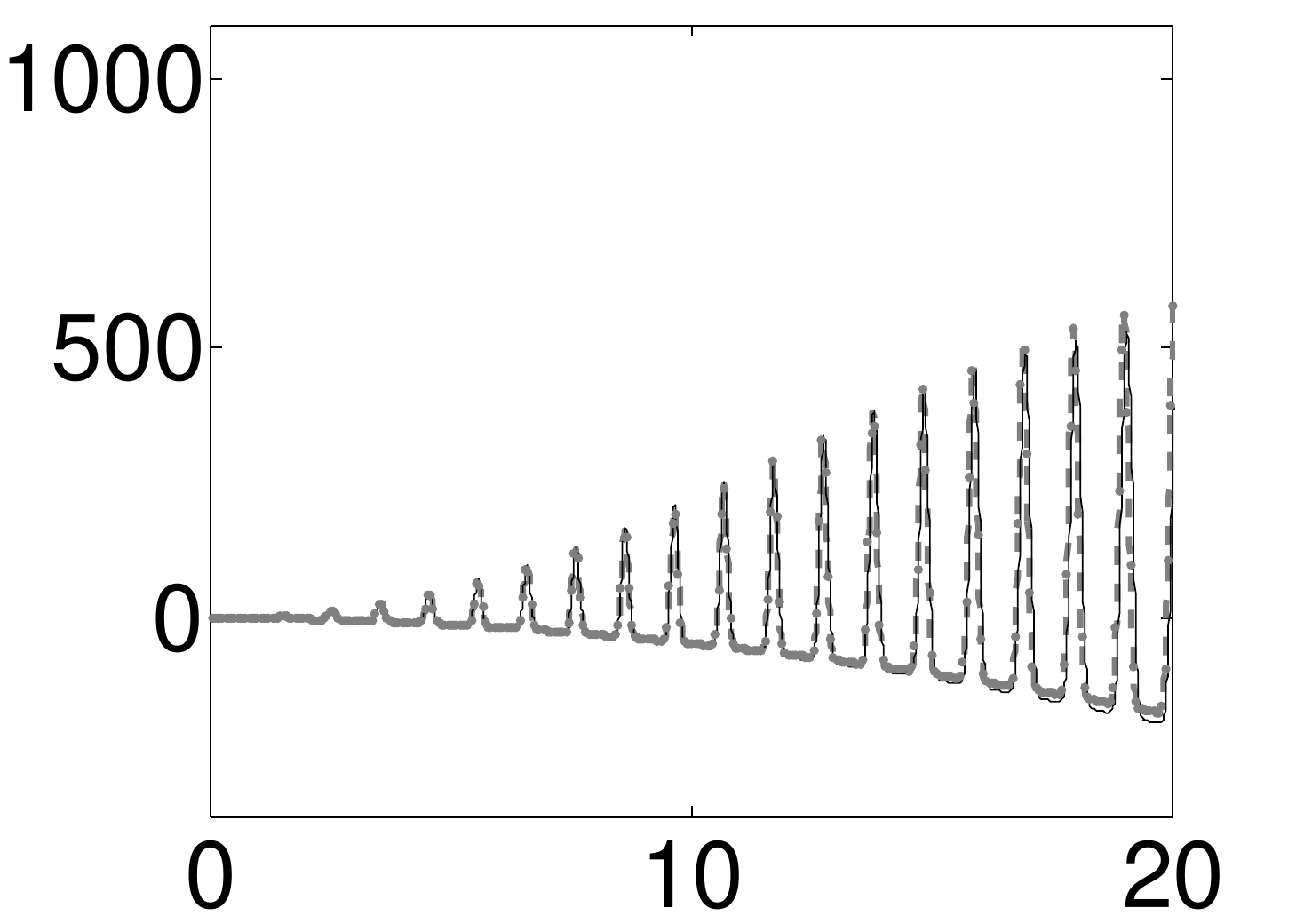}}%
    \put(1.14,2.37){\color[rgb]{0,0,0}\makebox(0,0)[lb]{ { $D=3,\alpha=2$} }}%        
    \put(0.91,2.1){\color[rgb]{0,0,0}\makebox(0,0)[lb]{ { $\delta g$} }}
    \put(1.35,1.7){\color[rgb]{0,0,0}\makebox(0,0)[lb]{ { $E/\hbar \omega$} }}%
    	\put(1.95,1.8){\includegraphics[width=1\unitlength]{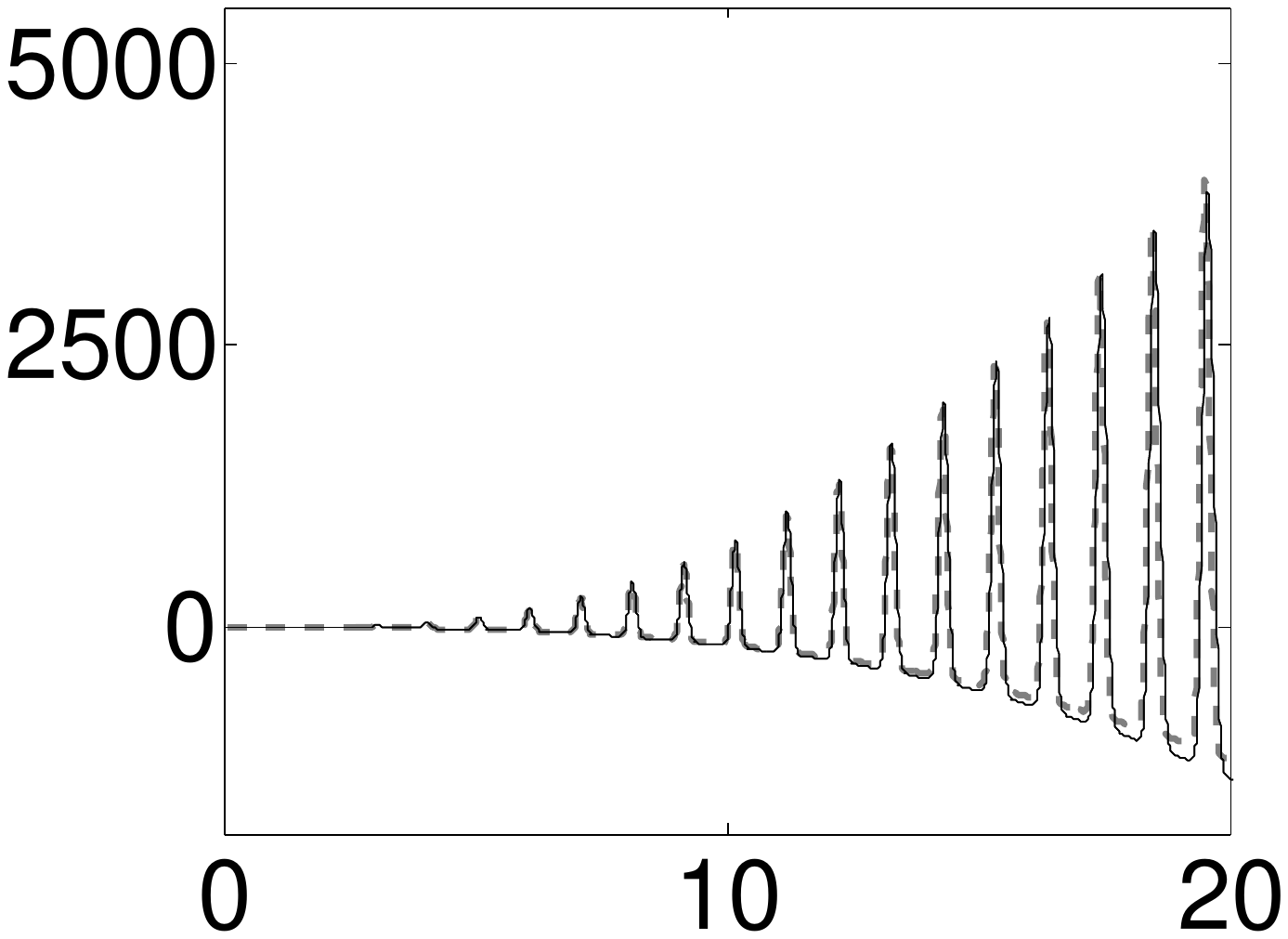}}%
    \put(2.17,2.37){\color[rgb]{0,0,0}\makebox(0,0)[lb]{ { $D=4,\alpha=2$} }}%
    \put(1.95,2.1){\color[rgb]{0,0,0}\makebox(0,0)[lb]{ { $\delta g$} }}
    \put(2.37,1.7){\color[rgb]{0,0,0}\makebox(0,0)[lb]{ { $E/\hbar \omega$} }}%    
%%%%%%%%%%%%%%%%%%%%%%%%%%%%%%%%
    	\put(-0.05,0.95){\includegraphics[width=1\unitlength]{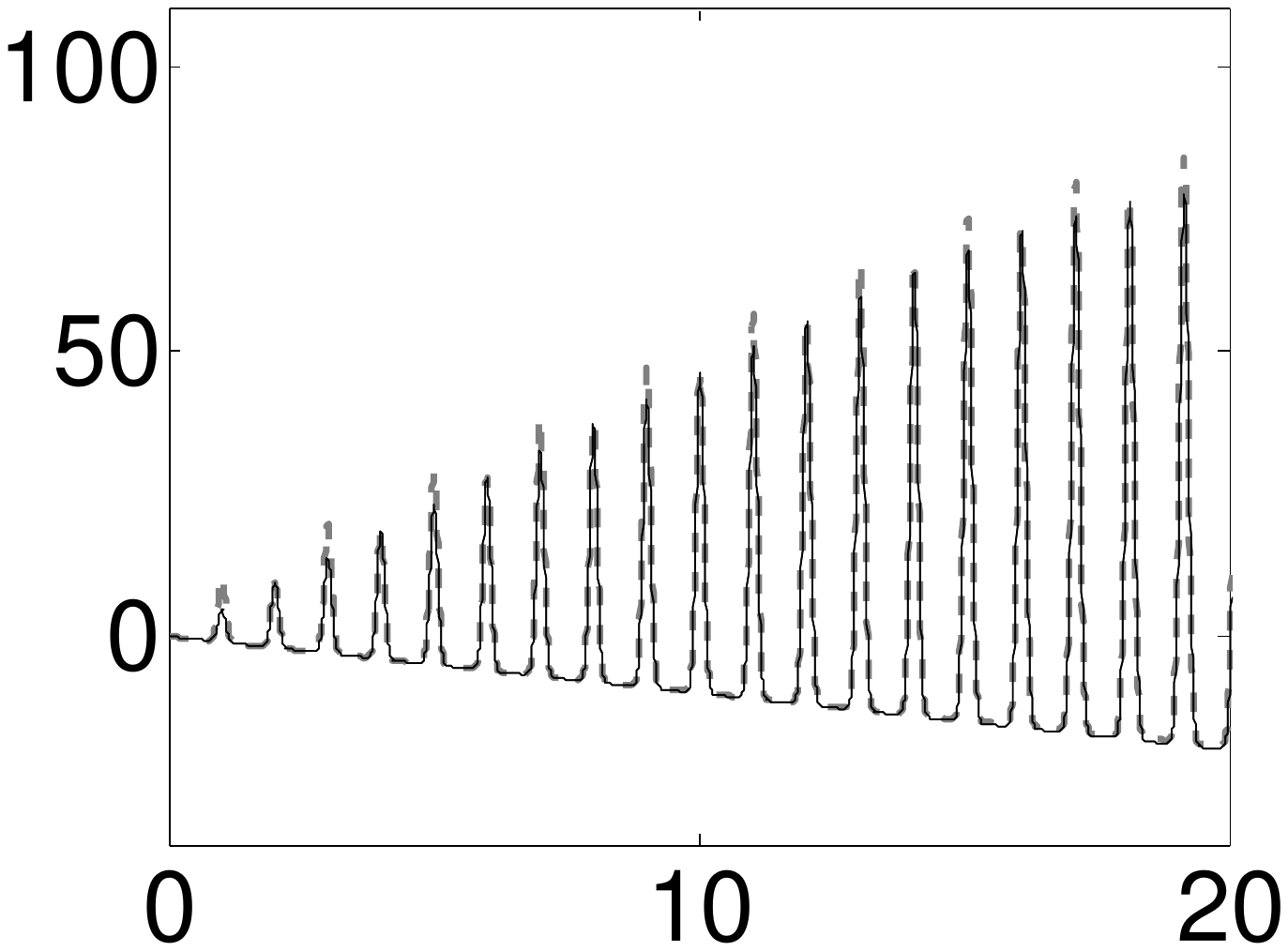}}%    	
    \put(0.1,1.52){\color[rgb]{0,0,0}\makebox(0,0)[lb]{ { $D=2,\alpha=3$} }}%    
    \put(-0.13,1.25){\color[rgb]{0,0,0}\makebox(0,0)[lb]{ { $\delta g$} }}
    \put(0.32,0.85){\color[rgb]{0,0,0}\makebox(0,0)[lb]{ { $E/\hbar \omega$} }}%
        \put(0.95,0.95){\includegraphics[width=1\unitlength]{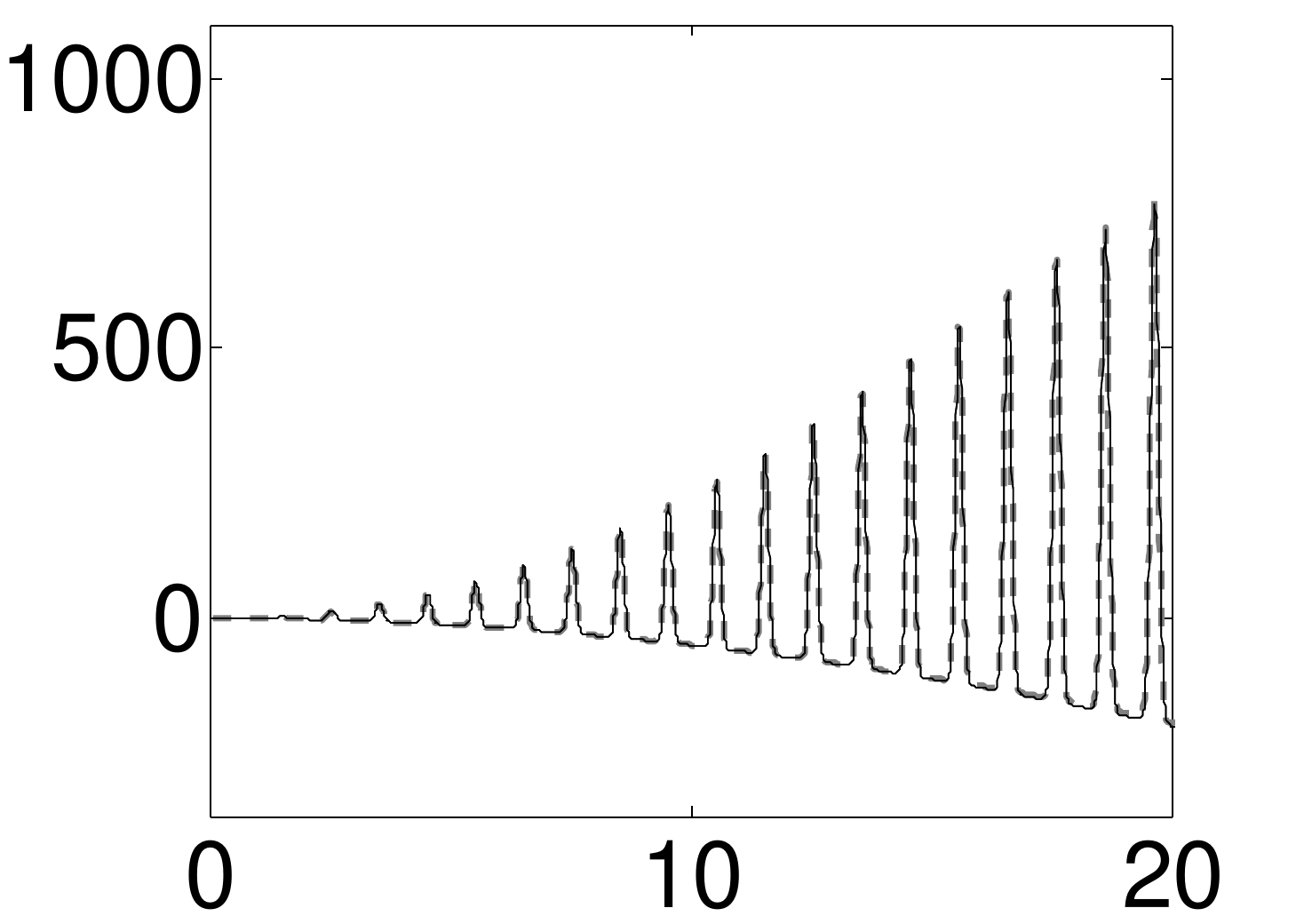}}%
    \put(1.14,1.52){\color[rgb]{0,0,0}\makebox(0,0)[lb]{ { $D=3,\alpha=3$} }}%
    \put(0.91,1.25){\color[rgb]{0,0,0}\makebox(0,0)[lb]{ { $\delta g$} }}
    \put(1.35,0.85){\color[rgb]{0,0,0}\makebox(0,0)[lb]{ { $E/\hbar \omega$} }}%
        \put(1.95,0.95){\includegraphics[width=1\unitlength]{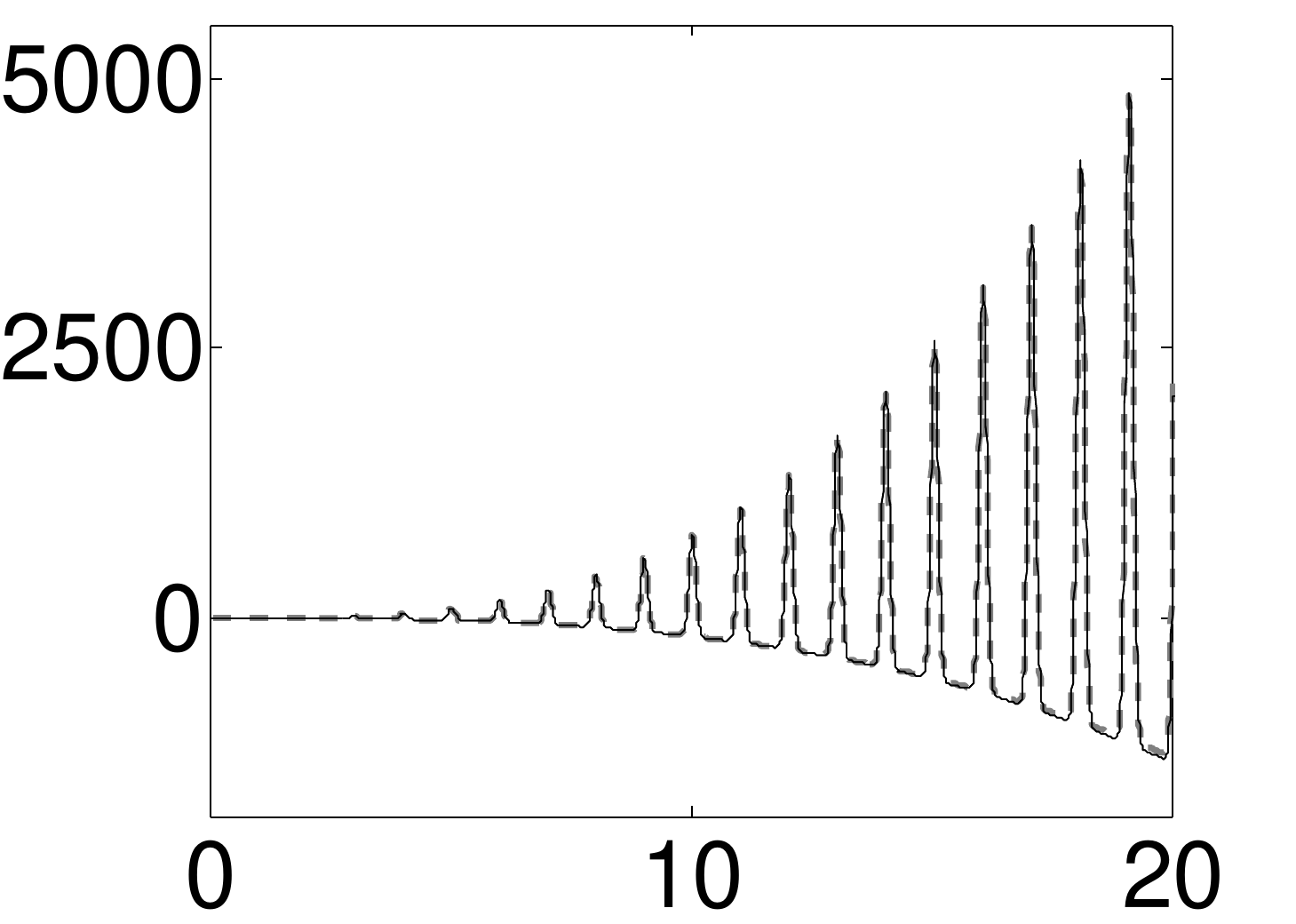}}%
    \put(2.17,1.52){\color[rgb]{0,0,0}\makebox(0,0)[lb]{ { $D=4,\alpha=3$} }}%
    \put(1.95,1.25){\color[rgb]{0,0,0}\makebox(0,0)[lb]{ { $\delta g$} }}
    \put(2.37,0.85){\color[rgb]{0,0,0}\makebox(0,0)[lb]{ { $E/\hbar \omega$} }}%    
%%%%%%%%%%%%%%%%%%%%%%%%%%%%%%%
        \put(-0.05,0.1){\includegraphics[width=1\unitlength]{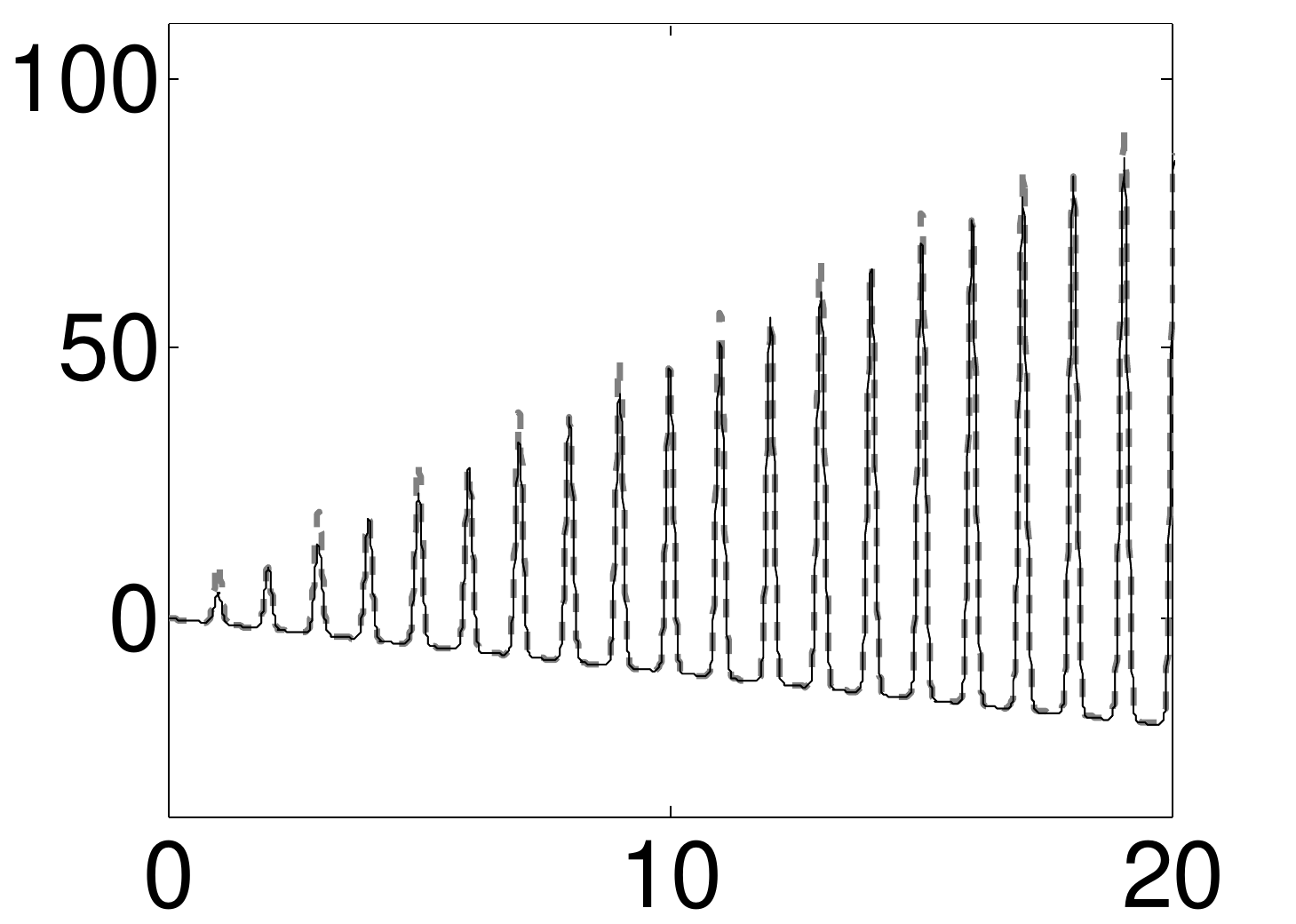}}%
    \put(0.1,0.67){\color[rgb]{0,0,0}\makebox(0,0)[lb]{ { $D=2,\alpha=4$} }}%
    \put(-0.13,0.4){\color[rgb]{0,0,0}\makebox(0,0)[lb]{ { $\delta g$} }}
    \put(0.32,0){\color[rgb]{0,0,0}\makebox(0,0)[lb]{ { $E/\hbar \omega$} }}%
        \put(0.95,0.1){\includegraphics[width=1\unitlength]{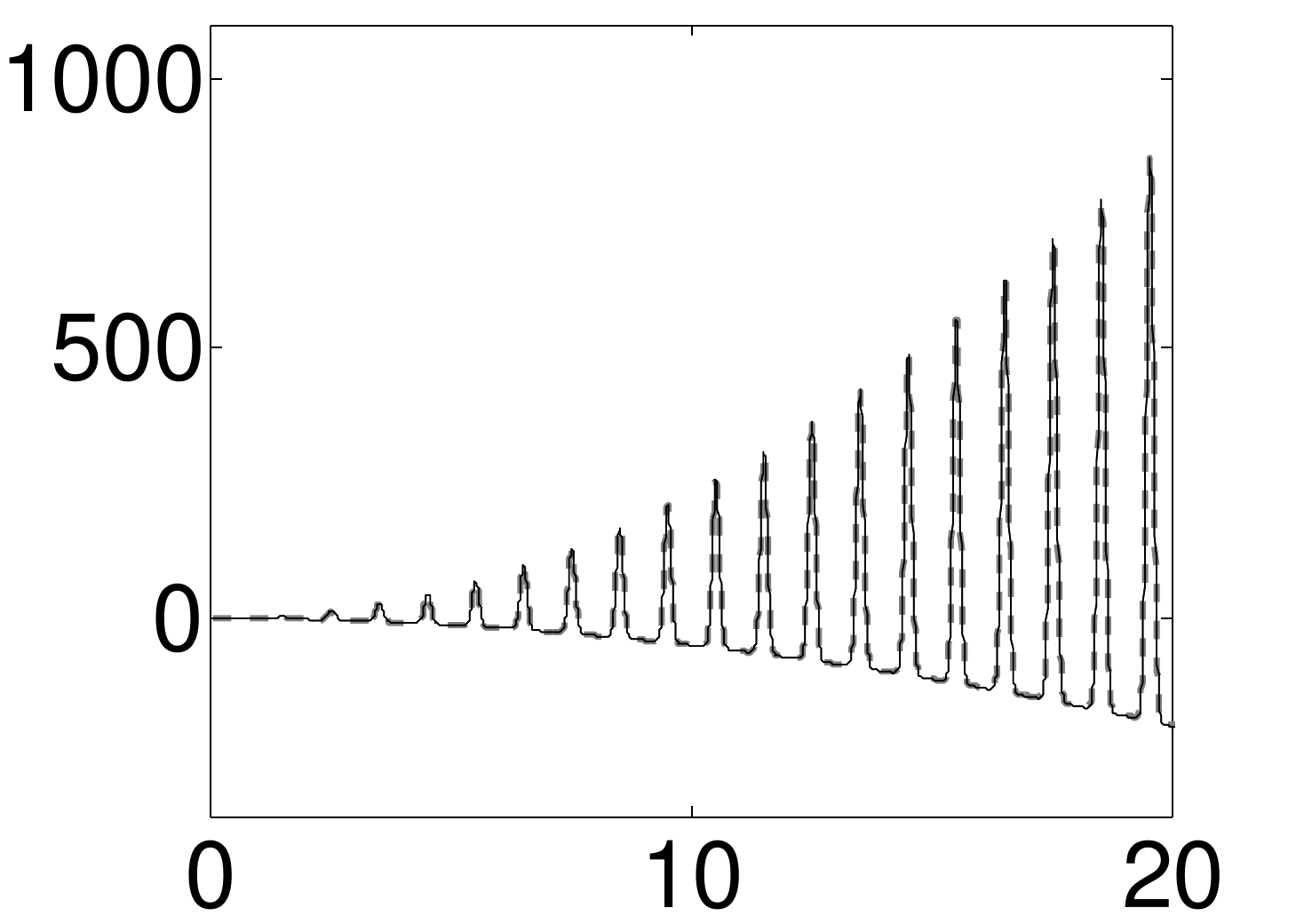}}%
    \put(1.14,0.67){\color[rgb]{0,0,0}\makebox(0,0)[lb]{ { $D=3,\alpha=4$} }}%
    \put(0.91,0.4){\color[rgb]{0,0,0}\makebox(0,0)[lb]{ { $\delta g$} }}
    \put(1.35,0){\color[rgb]{0,0,0}\makebox(0,0)[lb]{ { $E/\hbar \omega$} }}%
        \put(1.95,0.1){\includegraphics[width=1\unitlength]{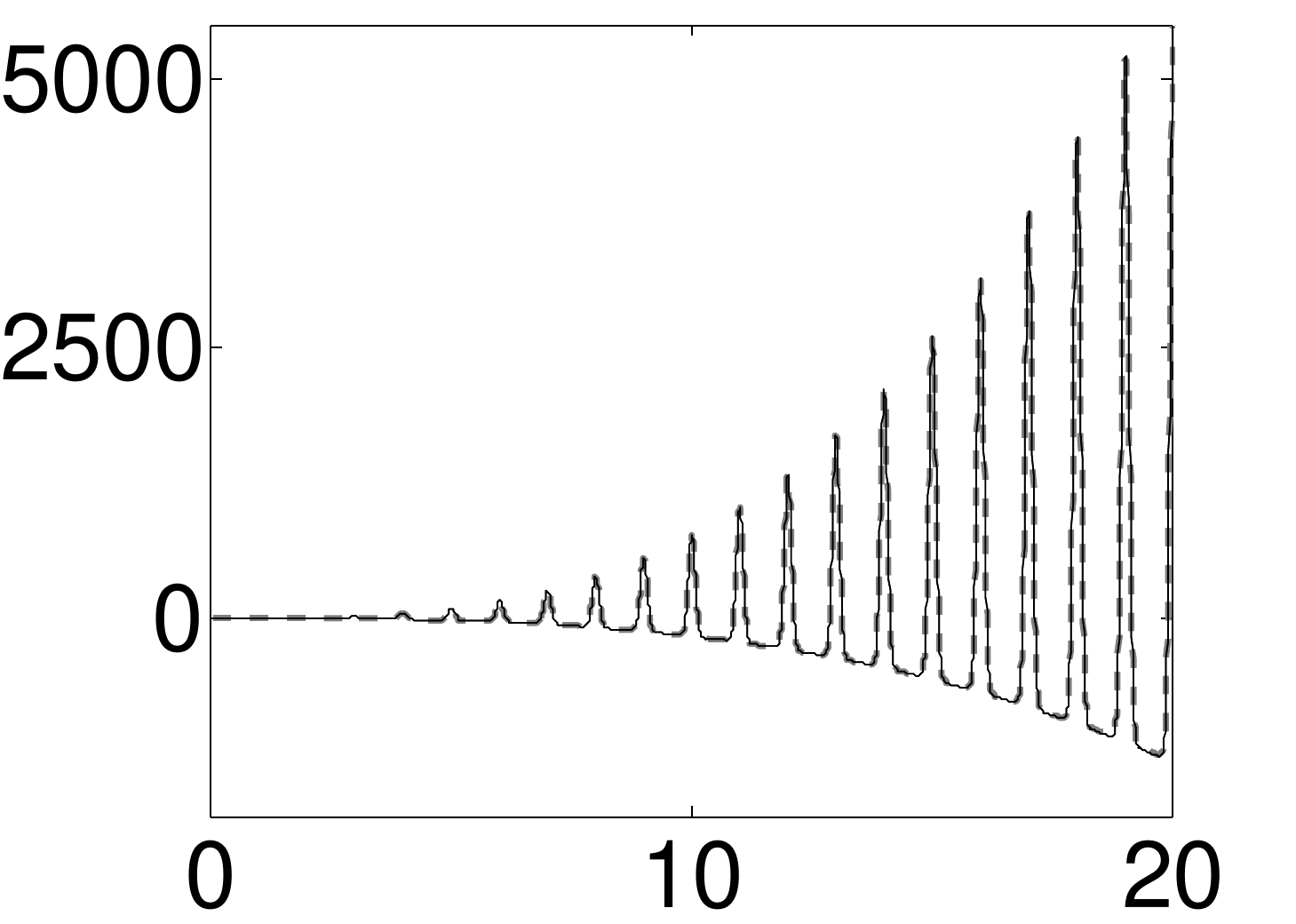}}%
    \put(2.17,0.67){\color[rgb]{0,0,0}\makebox(0,0)[lb]{ { $D=4,\alpha=4$} }}%
    \put(1.95,0.4){\color[rgb]{0,0,0}\makebox(0,0)[lb]{ { $\delta g$} }}
    \put(2.37,0){\color[rgb]{0,0,0}\makebox(0,0)[lb]{ { $E/\hbar \omega$} }}%
  \end{picture}%

\caption{Numerical comparison of the Gaussian-averaged perturbative trace formulae
and the DOS of the non-perturbative EBK spectra. The panels are oriented as in figure \ref{fig:SSS} with the same values for $\varepsilon$. Solid thin black curves shows the DOS of the
trace formulae (\ref{eq:convoluted_delta_g_pert}), while gray thick
dashed curves in the background shows the corresponding EBK results
obtained from (\ref{g_WKB}) and (\ref{dg_WKB}). For the case $D=3$ and $\alpha=2$ the Gaussian-averaged DOS
of a non-perturbative trace formula from \cite{JoPA2005} have also been
plotted with sparse gray dots. The Gaussian-averaged width is $w=0.1$
in all cases (physical parameters are set to unity). \label{fig:EBK} }
\end{figure}

As an alternative numerical semiclassical analysis for the monomial
potentials, we here briefly present results for the density of states
obtain from (non-perturbative) EBK energies $E_{n,l}\left(D,\alpha,\varepsilon,\omega\right)$
\cite{BBBook,CurtisEllis2004}. In order to perform a relevant numerical comparison with the perturbative
trace formulae for the gross structure of the DOS, we need to convolute the
EBK energies with a normalised Gaussian of width $w$ \cite{BBBook}
\begin{equation}
g_{\textnormal{EBK}}\left(E\right)=\frac{1}{w\sqrt{\pi}}\sum_{n,l}\frac{\left(2l+D-2\right)\left(l+D-3\right)!}{\left(D-2\right)! \: l!}e^{-\left[E-E_{n,l}\left(D,\alpha,\varepsilon,\omega\right)\right]^{2}/w^{2}}.\label{g_WKB}
\end{equation}
The pre-factor in the summand above gives the $l$-degeneracy for
radially symmetric systems, e.g., it is $2l+1$ for $D=3$. 
We can then
obtain the oscillating part of the EBK DOS, i.e. $\delta g_{\textnormal{EBK}}=g_{\textnormal{EBK}}-\bar{g}$,
by subtracting the Thomas-Fermi DOS of the potential in (\ref{eq:FullHamiltonian}).
For $\bar{g}$ we here use the expression for the classical orbits of length zero
introduced by Berry and Mount in \cite{bermo}, which is in the following
formulated for radially symmetric monomial perturbations to the HO
in $D$ spatial dimensions ($m=1$)
\begin{equation}
\bar{g}\left(E\right)=\left(2\pi\hbar^{2}\right)^{-D/2}\int_{0}^{r_{\mathrm{max}}}\frac{2\pi^{D/2}}{\left[\Gamma\left(D/2\right)\right]^{2}}\left[E-\frac{1}{2}\omega^{2}r^{2}-\varepsilon r^{2\alpha}\right]^{D/2-1}r^{D-1}dr.\label{dg_WKB}
\end{equation}
The upper integration limit $r_{\mathrm{max}}$ in (\ref{dg_WKB}) is the classical
turning point of the potential in (\ref{eq:FullHamiltonian}) and
hence is a real positive solution to the depressed polynomial
equation $r^{2\alpha}+\omega^{2}/\left(2\varepsilon\right)r^{2}-E/\varepsilon=0$.

In analogy with (\ref{g_WKB}), the perturbative trace formula (\ref{GeneralFinalTF})
is averaged with the same width $w$ according to \cite{BBBook}

\begin{equation}
\delta g_{\mathrm{pert}}(E)\simeq\frac{E^{D-1}}{\left(D-2\right)!\left(\hbar\omega\right)^{D}}\,\text{Re}\left\{ \sum_{k\neq0}(-1)^{Dk}e^{-\left[w kT_{0}/\left(2\hbar\right)\right]^{2}}\int_{0}^{1}\ell^{D-2}e^{-ik\sigma_{\alpha}\ell^{\alpha}P_{\alpha}(\frac{1}{\ell})/\hbar}d\ell\ e^{ikS_{0}/\hbar}\right\} ,\label{eq:convoluted_delta_g_pert}
\end{equation}
where the omitted $k=0$ term in (\ref{eq:convoluted_delta_g_pert})
corresponds to the substraction of the smooth Thomas-Fermi DOS. The
damping factor in (\ref{eq:convoluted_delta_g_pert}) suppress the
contributions of large $\left|k\right|$ terms and $T_{0}=2\pi/\omega$
is the period of the unperturbed HO.

In figure \ref{fig:EBK} we have plotted $\delta g_{\textnormal{EBK}}$
in addition to the results of the semiclassical trace formulae $\delta g_{\mathrm{pert}}$
for $\left|k\right|\leq10$ and they compare well for weak perturbations,
i.e. for small $\varepsilon$ and/or $E$. 
Qualitatively the results from our
comparisons agree with the investigation for the case $D=3$
and $\alpha=2$ undertaken in \cite{JoPA2005}, where it was found
that the beating pattern occur earlier for the perturbative trace
formula, see figure 3 in \cite{JoPA2005}. For the particular case
in \cite{JoPA2005} it was also derived a non-perturbative uniform
trace formula based on EBK theory. We have plotted the corresponding
result of this uniform trace formula with dots on top of the curve
for $\delta g_{\textnormal{EBK}}$ in the upper-mid-subfigure of figure
\ref{fig:EBK}.

\section{Summary}

As Bohr discovered 100 years ago, one can obtain information about a quantum system by study its classical counterpart.
We present a calculation of the gross structure of the quantum mechanical density of states in the form of a perturbative semiclassical trace formula.
We have generalised earlier work of Creagh \cite{crpert} and Brack \textit{et al.} \cite{JoPA2005}, in order to handle a $D$-dimensional harmonic oscillator perturbed by an arbitrary monomial potential.
The leading order perturbative classical action was found to be an even polynomial in a scaled angular momentum (\ref{eq:DeltaS_as_a_function_of_L}). 
These polynomials (table \ref{table_for_Delta_S}) are independent of the spatial dimension, 
and they have a simple representation (\ref{Delta_S_conjecture}) with help of the well known Legendre polynomials.  
Utilizing the equivalence between averaging the classical periodic orbits over a $n=2D-1$ dimensional sphere $S^n$, and a complex $n=D-1$ dimensional projective space $\mathbb{C}P^{n}$, we obtained the modulation factor for the perturbed trace formula.
This high dimensional integral (\ref{eq: ModulationFactorSphereIntegral}) was then reduced to a one-dimensional Fourier integral (\ref{OneDimensionalIntegral}).
For the two lowest orders of perturbative monomial potentials (e.g., coming from the leading order of Taylor expansions of more general potentials) the modulation factor was even calculated exactly (\ref{eq:HyperModulationfactor}). In odd dimensions, this modulation factor may be given in elementary functions (table  \ref{table_for_M_k}). 
In any dimension and perturbation, employing the stationary phase approximation, (\ref{SPA_final}) gives the leading order term of the modulation factor (figure \ref{fig:ModulationFactor}) which is sufficient for the perturbative periodic orbit theory presented.
In particular, this result can explain the occurrence of super-shell structures (figure \ref{fig:SSS}), seen earlier for the quartic perturbed three-dimensional ($D=3$) harmonic oscillator \cite{JoPA2005}. 
A prominant super-shell structure will occur for quartic- and sextic-perturbations ($\alpha=2,3$), when there are only two terms of the same order in $\hbar$ in the modulation factor.
In these cases the perturbative trace formula can be written in the form of only one sine-function for the slow envelope modulation, multiplied with one cosine-function for the fast beating modulation (\ref{SSS_TF}).

Our main results are that the classical diameter- and circular-periodic orbits are responsible for the gross quantum-shell structure for radially symmetric polynomial perturbations to the $D$-dimensional harmonic oscillator and that the resulting semiclassical trace formulae have been explicitly derived to leading order in $\hbar^{-1}$. Finally, the perturbative trace formulae have been numerically compared with non-perturbative EBK theory for small perturbations (figure \ref{fig:EBK}).

\section*{Acknowledgement}

We are grateful to colleagues and friends in Copenhagen for encouragement
to publish this work, and in particular to J. Gravesen and S. Markvorsen for reading an early version of the manuscript. 
We also thank N. Temme for expert advice on the end-point correction in (\ref{SPAI0}), N. Eriksen and P. Br\"{a}nd\'{e}n for discussions about polynomials, and J. Kvistholm for assistance with figures~\ref{figure1} and~\ref{figure2}. 
Finally we acknowledge comments from an anonymous referee that led to improvements in the presentation.


\begin{thebibliography}{10}

\bibitem{Bohr1913} N. Bohr: On the Constitution of Atoms and Molecules, \textit{Philos. Mag.} \textbf{26}, 1 (1913); \textit{ibid} p. 476.

\bibitem{Sommerfeld1916} A. Sommerfeld: Zur Quantentheorie der Spektrallinien, \textit{Annalen der Phys.} \textbf{51},
1 (1916).

\bibitem{e} A. Einstein: Zum Quantensatz von Sommerfeld und Epstein, \textit{Verh. Dtsch. Phys. Ges.} \textbf{19},
82 (1917).

\bibitem{b} L. Brillouin: Remarques sur la m\'{e}canique ondulatoire,  \textit{J. Phys. Radium} \textbf{7}, 353 (1926).

\bibitem{k} J. B. Keller: Corrected Bohr-Sommerfeld Quantum Conditions for Nonseparable Systems, \textit{Ann. Phys. (N. Y.)} \textbf{4}, 180 (1958).

\bibitem{VanVleck1928} J. H. van Vleck: The Correspondence Principle in the Statistical Interpretation of Quantum Mechanics, \textit{Proc. Natl. Acad. Sci USA}
\textbf{14}, 178 (1928).

\bibitem{Feynman1948} R. P. Feynman: Space-time approach to non-relativistic quantum mechanics, \textit{Rev. Mod. Phys.} \textbf{20},
367 (1948).

\bibitem{chaos-book1} M. C. Gutzwiller: \textit{Chaos in Classical and
Quantum Mechanics}, Springer Verlag, New York 1990.

\bibitem{chaos-book2} H. J. St\"{o}ckmann: \textit{Quantum Chaos: an Introduction}, Cambridge University Press,
Cambridge, UK 1999.

\bibitem{chaos-book3} F. Haake: \textit{Quantum Signatures of Chaos}, Springer, 2nd edition 2001.

\bibitem{gutz} M. C. Gutzwiller: Periodic Orbits and Classical Quantization Conditions, \textit{J. Math. Phys.} \textbf{12},
343 (1971), and references therein.

\bibitem{bablo} R. Balian and C. Bloch: Asymptotic evaluation of the Green's function for large quantum numbers,  \textit{Ann. Phys.} (N.Y.) \textbf{69},
76 (1972).

\bibitem{bermo} M. V. Berry and K. E. Mount: Semiclassical approximations in wave mechanics, \textit{Rep. Prog. Phys.} \textbf{35}, 315 (1972).

\bibitem{bertab} M. V. Berry and M. Tabor: Closed orbits and the regular bound spectrum, \textit{Proc. R. Soc. Lond. A}
\textbf{349}, 101 (1976).

\bibitem{Selberg1956} A. Selberg: Harmonic analysis and discontinuous groups in weakly symmetric Riemannian spaces with applications to Dirichlet series, \textit{J. Indian Math. Soc.} \textbf{20}, 47 (1956).

\bibitem{BBBook} M. Brack and R. K. Bhaduri: \textit{Semiclassical
Physics}, revised edition, Westview Press, Boulder, USA 2003.

\bibitem{struma} V. M. Strutinsky: Semi-classical theory of nuclear shell structure, \textit{Nukleonika} (Poland) \textbf{20}, 679 (1975).

\bibitem{struma2} V. M. Strutinsky and A. G. Magner: Quasiclassical theory of nuclear shell structure,  \textit{Sov. J. Part. Nucl.} \textbf{7}, 138 (1976).
 
\bibitem{nish} H. Nishioka, K. Hansen and B. R. Mottelson: Supershells in metal clusters, \textit{Phys. Rev. B}
\textbf{42}, 9377 (1990).

\bibitem{klavs} J. Pedersen, S. Bj{\o}rnholm, J. Borggren, K. Hansen , T. P. Martin and H. D. Rasmussen: Observation of quantum supershells in clusters of sodium atoms, \textit{Nature} \textbf{353}, 733 (1991).

\bibitem{wire} A. I. Yanson, I. K. Yanson and J. M. van Ruitenbeek: Observation of shell structure in sodium nanowires, \textit{Nature} \textbf{400}, 144 (1999).

\bibitem{wire2} A. I. Yanson, I. K. Yanson and J. M. van Ruitenbeek: Supershell Structure in Alkali Metal Nanowires, \textit{Phys. Rev. Lett.} \textbf{84},
5832 (2000).

\bibitem{YuPRA2005} Y. Yu, M. \"{O}gren, S. {\AA}berg, S. M. Reimann and M. Brack: Supershell structure in trapped dilute Fermi gases, \textit{Phys. Rev. A } \textbf{72}, 051602(R) (2005).

\bibitem{OgrenPRA2007} M. \"{O}gren and H. Heiselberg: Super-shell structures and pairing in ultracold trapped Fermi gases, \textit{Phys. Rev. A} \textbf{76}, 021601 (2007).

\bibitem{OlofssonPRL2008} H. Olofsson, S. {\AA}berg and P. Leboeuf: Semiclassical Theory of Bardeen-Cooper-Schrieffer Pairing-gap Fluctuations , \textit{Phys. Rev. Lett.} \textbf{100}, 037005 (2008).

\bibitem{AkolaPRB2008} J. Akola, H. P. Heiskanen and M. Manninen: Edge-dependent selection rules in magic triangular graphene flakes,\textit{ Phys. Rev. B} \textbf{77}, 193410 (2008).

\bibitem{crpert} S. C. Creagh: Trace Formula for Broken Symmetry, \textit{Ann. Phys.} (N. Y.) \textbf{248}, 60 (1996).

\bibitem{JoPA2005} M. Brack, M. \"{O}gren, Y. Yu and S. M. Reimann:  Uniform semiclassical trace formula for $U(3) \to SO(3)$ symmetry breaking, \textit{J. Phys. A} \textbf{38}, 9941 (2005); an extended version including
Appendices C and D can be found from: arXiv:nlin/0505060v2 {[}nlin.SI{]}.

\bibitem{BengtssonIJMPA2002} I. Bengtsson, J. Br\"{a}nnlund and K. \.{Z}yczkowski: $CP^n$, or, entanglement illustrated, \textit{Int. J. Mod. Phys. A} \textbf{17}, 4675 (2002).

\bibitem{ArnoldClassicalMechanics} V. I. Arnold: {\em Mathematical Methods of Classical Mechanics}, Springer. 1989.

\bibitem{Sakai} T. Sakai: {\em Riemannian Geometry}, Translations of Mathematical Monographs No. 149, American Mathematical Society. 1995.

\bibitem{Ogren} M. \"{O}gren: \emph{Shell structure and semiclassics,
Super-shells in trapped fermions}, verlag DM, Germany, 2009. An extended version of a dissertation at Lund University 2008.

\bibitem{GeometricMechanics} D. D. Holm, T. Schmah and C. Stocia: {\em Geometric Mechanics and Symmetry}, Oxford University Press,
New York 2009.

\bibitem{ONeill} B. O'Neill: {\em Semi-Riemannian Geometry, with Applications to Relativity}, Academic Press, 1983.

\bibitem{Bailey} W. N. Bailey: {\em Generalised Hypergeometric Series}, Cambridge University Press, England 1935.

\bibitem{wong} R. Wong: \textit{Asymptotic Approximation of Integrals}, Classics in Applied Mathematics, Vol. 34, SIAM, Philadelphia 2001.

\bibitem{HeiselbergPRL2002} H. Heiselberg and B. R. Mottelson: Shell Structure and Pairing for Interacting Fermions in a Trap, \textit{Phys. Rev. Lett.}
\textbf{88}, 190401 (2002).

\bibitem{AbramowitzStegun1972} M. Abramowitz and I. A. Stegun: {\em Handbook of Mathematical Functions}, Dover, New York 1972.

\bibitem{CurtisEllis2004}L. J. Curtis and D. G. Ellis: Use of the Einstein-Brillouin-Keller action quantization,
\textit{Am. J. Phys.} \textbf{72}, 1521 (2004).

\end{thebibliography}
\end{document}